\documentclass[journal=jacsat,manuscript=article]{achemso}
\usepackage{achemso}
\usepackage[version=3]{mhchem}
\usepackage[english]{babel}
\usepackage{filecontents}
\usepackage{graphicx}
\usepackage{courier}
\usepackage{amssymb,amsmath,amsthm,mathrsfs,comment,afterpage,accents}
\usepackage{mathtools}
\usepackage{braket}
\usepackage{mathrsfs}
\usepackage{courier}
\usepackage{color}
\usepackage{subdepth}
\usepackage{amsmath}
\setkeys{acs}{articletitle = true}
\makeatletter
\def\@dotsep{4.5}
\makeatother
\usepackage{dcolumn}
\usepackage{bm}
\renewcommand\vec\mathbf
\newcommand\mat\mathbf

\newcommand{\op}[1]{\hat{\mathcal{#1}}}
\setkeys{acs}{maxauthors = 0}
\newcommand{\insertnew}[1]{{\textcolor{black} {#1}}}
\newcommand{\replacewith}[2]{{\textcolor{red}{}}{\textcolor{black}{#2}}}
\newcommand{\revreplace}[2]{{\textcolor{black}{}}{\textcolor{black}{#2}}}
\newcommand{\revinsert}[1]{\textcolor{black}{#1}}

\title{
Open-Shell Coupled-Cluster Valence-Bond Theory Augmented with an Independent Amplitude Approximation for Three-Pair Correlations:
Application to a Model
Oxygen-Evolving Complex and 
Single Molecular Magnet
}
\author{Joonho Lee}
\email{linusjoonho@gmail.com}
\author{David W. Small}
\email{dsmallchem@gmail.com }
\author{Martin Head-Gordon}
\email{mhg@cchem.berkeley.edu}
\affiliation{
Department of Chemistry, University of California, Berkeley, California 94720, USA
Chemical Sciences Division, Lawrence Berkeley National Laboratory, Berkeley, California 94720, USA
}
\begin{document}
\maketitle
\newpage
\begin{abstract}
We report the failure of coupled-cluster valence-bond (CCVB) theory with two-pair configurations [{\it J. Chem. Phys.} {\textbf {2009}}, {\it 130}, 084103 (2009)] for \insertnew{open-shell (OS)} spin-frustrated systems where including three-pair configurations is necessary to properly describe strong spin-correlations. We extend \insertnew{OS-}CCVB by augmenting the model with three-pair configurations within the independent amplitude approximation (IAA). The resulting new electronic structure model, \insertnew{OS-}CCVB+i3, involves only a quadratic number of independent wavefunction parameters. \insertnew{It includes the recently reported closed-shell CCVB+i3 as a special case.} Its cost is dominated by integral transformations and it is capable of breaking multiple bonds exactly for \replacewith{the}{all} systems examined so far. The strength of \insertnew{OS-}CCVB+i3 is highlighted in realistic systems including the [\ce{CaMn3O4}] cubane subunit of the oxygen-evolving complex and a molecular magnet with the [\ce{Cr9}] core unit as well as model systems such as \ce{N3}, \ce{V3O3}, and \ce{P5}. We show that \insertnew{OS-}CCVB+i3 is only slightly dependent on the underlying perfect-pairing reference while \insertnew{OS-}CCVB shows a stronger dependence. We also emphasize the compactness of the \insertnew{OS-}CCVB+i3 wavefunction compared to the heat-bath configuration interaction wavefunction, a recently introduced \insertnew{soft} exponential-scaling approach.
\end{abstract}
\newpage
\section{Introduction}
%

\insertnew{Low-order M{\o}ller-Plesset perturbation theory, based on the best possible independent particle model, qualitatively fails for systems with strong correlation (SC). 
Such failure is commonly observed in the bond dissociation of molecules.
Upon dissociating bonds, all the electrons involved become perfectly localized and the different spin states become all degenerate.
We define this particular type of strong correlation as strong spin-correlation (SSC).\cite{Small2011} 
SSC is often characterized by high energy costs for charge transfer excitations, versus very small energy costs for spin-flipping excitations that leave local charges unchanged. As a result SSC problems typically have wavefunctions in which the amplitudes for spin-flipping excitations are large and essential, while the amplitudes for charge transfer substitutions are small and of secondary importance.
}

There are a number of numerical techniques that can properly describe SSC and we mention some of the significant developments.
We first mention density matrix renormalization group (DMRG) theory by White \cite{White1992,White1993,White1999,Schollwock2011a} which was originally developed to simulate one-dimensional (1D) quantum lattice models. DMRG correctly encodes the entanglement area law \cite{Schollwock2011a} for 1D gapped systems with local Hamiltonian, and this is the key to its success in solving 1D problems. This technique from the condensed matter physics community has been successfully applied to quantum chemical problems \cite{Chan2011,Kurashige2013,Sharma2014} although the scaling is still exponential in higher dimensions than 1D. \revreplace{as well as for systems with long-range interactions.}{}

There are two promising quantum Monte Carlo (QMC) algorithms that work with Slater determinants. First, we mention Alavi and co-worker's full configuration interaction QMC (FCIQMC),\cite{Booth2009} which is formally exponential scaling. It, however, has \insertnew{significantly} pushed \insertnew{back} the \insertnew{onset} of the exponential wall to roughly 50 electrons. Sampling determinants stochastically, FCIQMC avoids the usual fixed-node approximations commonly used in diffusion MC. It has been applied to various molecular systems along with solid-state applications.\cite{Booth2013,Thomas2015} \revreplace{The scope of this method in quantum chemistry is currently comparable to that of DMRG.}{}

Another alternative is auxiliary field QMC (AFQMC).\cite{Sugiyama1986} AFQMC utilizes the Hubbard-Stratonovich transformation to elegantly cast an interacting many-body problem to non-interacting problems with a set of random auxiliary fields. Sampling an infinite number of auxiliary fields in principle converges to the exact answer. However, employing either a constrained-path or phaseless approximation is almost necessary in practice to control the sign problem in large systems at the expense of introducing bias.\cite{Zhang1995,Suewattana2007} Its application to chemical systems has been somewhat limited although preliminary results are promising.\cite{Al-Saidi2007,Purwanto2008,Purwanto2013,Purwanto2015,Purwanto2016,Shee2017}

From the quantum chemistry community, there are numerous brute-force approaches based on configuration interaction (CI) methods.\cite{Szalay2012} 
Most of them are in general exponential scaling \insertnew{
using configuration selection with second-order perturbation theory to reduce the prefactor relative to full CI. \cite{Bender1969,Langhoff1973,Huron1973,Buenker1974,Buenker1978,Evangelisti1983,Harrison1991,Caballol1992,Giner2013,Evangelista2014,Tubman2016,Holmes2016,Sharma2017,Smith2017,Holmes2017}}
A recently introduced, exponential-scaling heat-bath CI (HCI) also belongs to this category, \cite{Holmes2016,Sharma2017,Smith2017,Holmes2017} and this method is used for the benchmark purpose in this paper. 
To best of our knowledge, none of the methods in this category can exactly dissociate multiple bonds with only a polynomial amount of work with respect to the number of bonds.

Other quantum chemistry methods are mainly based on coupled-cluster (CC) approaches. It is a particularly promising direction as those methods generally involve only a polynomial number of wavefunction parameters to describe an exponential of number of configurations through the non-linear wavefunction \revreplace{ans{\"a}tz}{ansatz}. There are numerous approaches in this category,\cite{Piecuch1992,Piecuch1990b,Jankowski1991,Piecuch1993,VanVoorhis2000a,Piecuch2002,Piecuch2004,Li2004,Piecuch2005,Krylov2006,Bartlett2007,Small2009,Parkhill2009,Parkhill2010,Piecuch2010,Huntington2010,Malrieu2010,Lyakh2010,Small2011,Xu2011,Robinson2011,Evangelista2011,Lyakh2012,Small2012,Kats2013,Small2014,Small2017,Henderson2017} and our method discussed below also falls into this. Interested readers are referred to Introduction of ref. \citenum{Lee2017} and the references therein.

Our group has been developing a powerful polynomial-scaling approach to bond-breaking based on CC valence bond (CCVB).\cite{Small2009,Small2011,Small2012,Small2014,Lee2017,Small2017}
It encodes strong spin-correlations by excitations from a generalized valence bond perfect pairing (GVB-PP) reference \cite{Goddard1971,Goddard1973} and involves only a quadratic number of wavefunction parameters associated with two-pair (2P) substitutions. Furthermore, it yields a spin-pure, size-extensive wavefunction and its cost is dominated by integral transformation as long as the amplitude equation is solved by a computationally inexpensive way. The CCVB \revreplace{ans{\"a}tz}{ansatz} generalizes spin-projected unrestricted Hartree-Fock (SPUHF) \cite{Lowdin1955,Mayer1973,Nakatsuji1973,Yamaguchi1978,Jimenez-Hoyos2012} to a size-extensive wavefunction at the expense of orbital-invariance and variationality. It can reach the correct asymptote when breaking bonds as long as UHF can properly dissociate. Moreover, in some cases CCVB can break bonds when UHF cannot reach the correct asymptote such as a triplet \ce{O2} dissociation to two triplet oxygens.

CCVB can be understood from various different perspectives. One of them is to look at CCVB from a VB perspective. The spin-coupled VB (SCVB) approach \cite{Gerratt1980,Cooper1991,Gerratt1997,Hiberty2007} can describe bond-breaking exactly within a given active space at an exponential-scaling cost. Applying a modified CC expansion with double excitations along with strong orthogonality between pairs, we obtain CCVB that is polynomial-scaling and practically identical to SCVB at dissociation limits. 

Another viewpoint is to start from CCVB with singles and doubles (CCVB-SD) \cite{Small2012, Lee2017} which is a full singles and doubles model like restricted CC singles and doubles (RCCSD) and parametrizes connected quadruples in a different way than RCCSD. Replacing singles with orbital optimization and applying the pairing active space constraint and the local approximation (i.e., sparsifying $T$-amplitudes), we obtain CCVB with an RHF reference. One could then write the same wavefunction with a GVB-PP reference by converting the GVB-PP amplitudes to GVB-PP polarization angles. This allows for writing the CCVB \revreplace{ans{\"a}tz}{ansatz} in the originally proposed form.

The goal of this paper is to demonstrate the failure of the original CCVB model for \insertnew{open-shell} spin-frustrated systems and the necessity to incorporate 3-pair (3P) substitutions in such systems. We also introduce an improved wavefunction where the number of independent wavefunction parameters scales still quadratically with system size, but it includes the influence of 3P substitutions within the independent amplitude approximation (IAA). We denote this new model as CCVB+i3 and we will present \insertnew{full} details later in the paper.  

This paper is organized as follows: we first review CCVB and then discuss the full 3P extension of this model, CCVB-3. We formally analyze CCVB-3 and discuss a subtle issue regarding the size-consistency of the model. We then introduce a new model, CCVB+i3 which is an attempt to include the 3P substitutions in a simpler, size-consistent manner. 
Lastly, we discuss interesting spin-frustrated model systems along with \insertnew{models of two} chemically relevant, realistic systems. Therein, we show promising results of CCVB+i3 compared to HCI.

\section{Theory}
\subsection{Notation}
In this paper we use $K,L,M,P,Q,R,\cdot\cdot\cdot$ to denote closed-shell (CS) pairs, $\mu,\nu,\lambda,\cdot\cdot\cdot$ to denote singly occupied orbitals, and $a, b, c,\cdot\cdot\cdot$ to denote either of them. More precisely, this means
\begin{align}
&1 \le K,L,M,P,Q,R, \cdot\cdot\cdot\le n_\beta\\
&n_\beta < \mu,\nu,\lambda, \cdot\cdot\cdot\le n_\alpha\\
&1 \le a,b,c, \cdot\cdot\cdot\le n_\alpha
\end{align}
where
$n_\alpha$ and $n_\beta$ denote the number of $\alpha$ and $\beta$ electrons, respectively.

We also establish a notation for several quantities which will be used throughout this paper.
The GVB-PP (or PP for short) reference is defined as
\begin{equation}
\Ket{\psi_0} = \prod_\mu \hat{a}_{\mu_\alpha}^\dagger\prod_{\substack{K}} \hat{g}_{s,K}^\dagger \Ket{0}
\end{equation}
where we used $\Ket{0}$ to denote the vacuum state, $\hat{a}_{\mu_\alpha}^\dagger$ is the fermionic creation operator, 
and the singlet pair (or geminal) $K$ creation operator $\hat{g}_{s,K}^\dagger$ is defined as
\begin{equation}
\hat{g}_{s,K}^\dagger = \frac{1}{\sqrt{2\left(1+\cos^2\theta_K\right)}}
\left(
2\cos\theta_K\hat{a}_{K_\alpha}^\dagger \hat{a}_{K_\beta}^\dagger
-\sin\theta_K\hat{a}_{K_\alpha}^\dagger \hat{a}_{\hat{K}_\beta}^\dagger
-\sin\theta_K\hat{a}_{\hat{K}_\alpha}^\dagger \hat{a}_{K_\beta}^\dagger
\right)
\end{equation}
where $\theta_K$ is the polarization angle for a pair $K$ and $K_\alpha$, $K_\beta$, $\hat{K}_\alpha$, and $\hat{K}_\beta$ denote the four spin-orbitals associated with the pair. $\theta_K = \pi/2$ corresponds to a fully polarized pair (i.e., a perfect diradical). Those pairs are strongly orthogonal which simplifies the calculation of matrix elements discussed later.

In CCVB, other configurations in addition to the PP reference are defined with excitations from the reference.
The 2P substitutions include 
a CS-CS substitution ($\delta_{s2}$),
\begin{equation}
\Ket{\psi_{(KL)}} = \hat{d}_{s2, KL}^\dagger \hat{g}_{s,K} \hat{g}_{s,L} \Ket{\psi_0},
\end{equation}
and 
a CS-OS substitution ($\delta_{d2}$),
\begin{equation}
\Ket{\psi_{(K\mu)}} = 
\hat{d}_{d2,K\mu}^\dagger
\hat{a}_{\mu_\alpha} \hat{g}_{s,K}
\Ket{\psi_0}
\end{equation}
where
\begin{equation}
\hat{d}_{s2, KL}^\dagger = 
\frac{1}{\sqrt{3}}
\left(
\hat{g}_{t1,K}^\dagger \hat{g}_{t1,L}^\dagger
-\hat{g}_{t2,K}^\dagger \hat{g}_{t3,L}^\dagger
-\hat{g}_{t3,K}^\dagger \hat{g}_{t2,L}^\dagger
\right),
\end{equation}
and
\begin{equation}
\hat{d}_{d2,K\mu}^\dagger = 
\frac{1}{\sqrt{3}}
\left(
\hat{g}_{t_1,K}^\dagger \hat{a}_{\mu_\alpha}^\dagger
+\sqrt{2}
\hat{g}_{t_2,K}^\dagger \hat{a}_{\mu_\beta}^\dagger
\right)
\end{equation}
with the triplet pair creation operators,
\begin{align}
\hat{g}_{t1,K}^\dagger &= \frac{1}{\sqrt{2}}
\left(
-\hat{a}_{K_\alpha}^\dagger \hat{a}_{\hat{K}_\beta}^\dagger
+\hat{a}_{\hat{K}_\alpha}^\dagger \hat{a}_{K_\beta}^\dagger
\right),\\
\hat{g}_{t2,K}^\dagger &= 
\hat{a}_{K_\alpha}^\dagger \hat{a}_{\hat{K}_\alpha}^\dagger,\\
\hat{g}_{t3,K}^\dagger &= 
\hat{a}_{K_\beta}^\dagger \hat{a}_{\hat{K}_\beta}^\dagger.
\end{align}
In a simpler term, $\delta_{s2}$ represents a substitution of a product of two singlet pairs with two triplet pairs coupled into a four-electron singlet configuration. Similarly, $\delta_{d2}$ is a substitution of a product of a singlet geminal with an alpha electron with a triplet pair and a unpaired electron coupled into a three-electron doublet configuration.

The 3P substitutions include a CS-CS-CS substitution ($\epsilon_{s5}$),
\begin{equation}
\Ket{\psi_{(KLM)}} = 
\hat{e}_{s5, KLM}^\dagger 
\hat{g}_{s,K} \hat{g}_{s,L} \hat{g}_{s,M} 
\Ket{\psi_0}
\end{equation}
and a CS-CS-OS substitution ($\epsilon_{d5}$),
\begin{equation}
\Ket{\psi_{(KL\mu)}} = \hat{e}_{d5, KL\mu}^\dagger \:\hat{g}_{s,K} \hat{g}_{s,L} \hat{a}_{\mu_\alpha} \Ket{\psi_0} 
\end{equation}
where
\begin{align}\nonumber
\hat{e}_{s5, KLM}^\dagger = 
\frac{1}{\sqrt{6}}
&(
\hat{g}_{t1,K}^\dagger \hat{g}_{t2,L}^\dagger \hat{g}_{t3,M}^\dagger
-\hat{g}_{t1,K}^\dagger \hat{g}_{t3,L}^\dagger \hat{g}_{t2,M}^\dagger
-\hat{g}_{t2,K}^\dagger \hat{g}_{t1,L}^\dagger \hat{g}_{t3,M}^\dagger\\
&+\hat{g}_{t2,K}^\dagger \hat{g}_{t3,L}^\dagger \hat{g}_{t1,M}^\dagger
+\hat{g}_{t3,K}^\dagger \hat{g}_{t1,L}^\dagger \hat{g}_{t2,M}^\dagger
-\hat{g}_{t3,K}^\dagger \hat{g}_{t2,L}^\dagger \hat{g}_{t1,M}^\dagger
),
\end{align}
and
\begin{equation}
\hat{e}_{d5, KL\mu}^\dagger = 
\frac{1}{\sqrt{6}}
\left(
-\sqrt{2}
\left(
\hat{g}_{t1,K}^\dagger \hat{g}_{t2,L}^\dagger
-\hat{g}_{t2,K}^\dagger \hat{g}_{t1,L}^\dagger
\right)\hat{a}_{\mu_\beta}^\dagger
+
\left(
\hat{g}_{t2,K}^\dagger \hat{g}_{t3,L}^\dagger
-\hat{g}_{t3,K}^\dagger \hat{g}_{t2,L}^\dagger
\right)\hat{a}_{\mu_\alpha}^\dagger
\right).
\end{equation}
Similarly to the 2P substitutions, $\epsilon_{s5}$ denotes a substitution of a product of three singlet pairs with a product of three triplet pairs coupled to an overall singlet and $\epsilon_{d5}$ represents a substitution of a product of two singlet pairs and an alpha electron with a product of two triplet pairs and a unpaired electron coupled to an overall doublet.

The higher-order substitutions are trivially defined with the definitions above by the virtue of a CC expansion. For instance, we have
\begin{equation}
\Ket{\psi_{(K\mu)(LMN)}} 
=
\hat{d}_{d2,K\mu}^\dagger
\hat{a}_{\mu_\alpha} \hat{g}_{s,K}
\Ket{\psi_{(LMN)}}
= 
\hat{d}_{d2,K\mu}^\dagger
\hat{a}_{\mu_\alpha} \hat{g}_{s,K}
\hat{e}_{s5, LMN}^\dagger 
\hat{g}_{s,L} \hat{g}_{s,M} \hat{g}_{s,N} 
\Ket{\psi_0}.
\end{equation}

Those substitutions are not necessarily orthogonal and generally linearly dependent for a given excitation level. We introduce a dual frame to $\{|\psi_i\rangle\}$, which we shall write $\{|\phi_i\rangle\}$. In particular, we define $\{|\phi_i\rangle\}$ to be the {\it canonical} dual frame of $\{|\psi_i\rangle\}$,
\begin{equation}
|\phi_i\rangle = \sum_{j}\left({\mathbf{S}^{+}}\right)_{ij}|\psi_j\rangle
\end{equation}
where $S_{ij} = \langle\psi_i|\psi_j\rangle$ and $\mathbf S^+$ is the pseudoinverse of $\mathbf S$.
Forming $\mathbf S^+$ can be done quite cheaply exploiting the block structure of $\mathbf S$. Moreover, in some special cases we have $|\phi_i\rangle$ = $|\psi_i\rangle$ as the relevant block in $\mathbf S$ forms an identity block. These include $|\phi_0\rangle=|\psi_0\rangle$, $|\phi_{(KL)}\rangle = |\psi_{(KL)}\rangle$, and $|\phi_{(KLM)}\rangle = |\psi_{(KLM)}\rangle$. Other than those three special cases, $|\phi_i\rangle$ is expected to be different from $|\psi_i\rangle$.
\subsection{Review of CCVB}
We review the CCVB wavefunction \revreplace{ans{\"a}tz}{ansatz} that includes only 2P substitutions (i.e., $(KL)$, $(K\mu)$) and disconnected higher-order substitutions arising from those.
The pertinent CC expansion in terms of $\{|\phi_i\rangle\}$ reads
\begin{align}
\label{eq:ccvb2p}
\Ket{\psi_\text{2P}}\nonumber
=& \Ket{\phi_0}  
+ \sum_{\substack{KL\\K<L}}t_{KL}\Ket{\phi_{(KL)}} + \sum_{\substack{K\mu}}t_{K\mu}\Ket{\phi_{(K\mu)}} \\ \nonumber
&+ \sum_{\substack{KLMN\\K<L<M<N}}
\left(
t_{KL}t_{MN}\Ket{\phi_{(KL)(MN)}}
+ t_{KM}t_{LN}\Ket{\phi_{(KM)(LN)}}
+t_{KN}t_{LM}\Ket{\phi_{(KN)(LM)}}
\right)\\ \nonumber
&+ \sum_{\substack{KLM\mu\\K<L<M}}\left(
t_{KL}t_{M\mu}\Ket{\phi_{(KL)(M\mu)}}
+ t_{KM}t_{L\mu}\Ket{\phi_{(KM)(L\mu)}}
+t_{K\mu}t_{LM}\Ket{\phi_{(K\mu)(LM)}}\right)\\
&+ \sum_{\substack{KL\mu\lambda\\K<L\\\mu<\lambda}}
\left(
t_{K\mu}t_{L\lambda} \Ket{\phi_{(K\mu)(L\lambda)}}
+t_{K\lambda}t_{L\mu}\Ket{\phi_{(K\lambda)(L\mu)}}
\right)
+\cdot\cdot\cdot,
\end{align}
where we listed only those terms that are necessary to solve the CCVB 2P amplitude equation.
The CCVB energy and the 2P amplitudes are computed via projection equations similar to those of regular CC methods:
\begin{align}\label{eq:ccvbenergy}
E&\equiv
\Bra{\psi_0}\op{H}\Ket{\psi_\text{2P}}
\\
E
t_{Ka}
&=
\Bra{\psi_{(Ka)}}\op{H}\Ket{\psi_\text{2P}}.
\end{align}
The 2P amplitude residual of CCVB reads
\begin{equation}
R_{Ka} = 
\Bra{\psi_{(Ka)}}\op{H}\Ket{\psi_\text{2P}} 
-
E
t_{Ka}.
\label{eq:T2}
\end{equation} 
When solving $R_{K\mu} = 0$, a complication arises
as $\Braket{\psi_{(K\mu)}|\psi_{(K\lambda)}} \ne \delta_{\mu\lambda}$ and $\Braket{\psi_{(KL)}|\psi_{(K\mu)(L\lambda)}}\ne 0$. In other words, we need to compute the pseudoinverse of a block in $\mathbf S$ that is not as small as that of the CS case. In particular, the size of the block of $\mathbf S$ that we need to pseudo-invert is now system-dependent.

To circumvent this complication, Small and Head-Gordon employed a supersystem approach which adds fictitious $\beta$ electrons to a high-spin original system to make it overall a CS system.\cite{Small2017}
In this approach, each OS $\alpha$ electron is coupled with a fictitious $\beta$ electron and they behave as a CS pair together in $\Ket{\Psi_0}$. We refer \insertnew{to} this pair \insertnew{composed} of an original system electron and a fictitious electron as an ``OS'' pair.
Note that we are using $\Ket{\Psi_0}$ to denote the closed-shell supersystem in contrast to $\Ket{\psi_0}$ which we used to denote the original system.
We work with a CCVB wavefunction of this fictitious supersystem whose complete set of spin configurations (which include all 3P substitutions as well) contain those of the original CCVB configurations.
Solving the modified 2P amplitude equation of the supersystem with some constraints is equivalent to solving the original CCVB amplitude equation. 

We seek $t_{Ka}$ that satisfies $\Omega_{Ka} = 0$
where
\begin{align}\label{eq:T2amp1}
\Omega_{KL} &= R_{KL},\\\label{eq:T2amp2}
\Omega_{K\mu} &= R_{K\mu} 
+ \sum_{\lambda\ne \mu} \kappa_{K\mu;\lambda} t_{K\mu\lambda},
\end{align}
where
\begin{equation}
\kappa_{ab;c} \equiv \Bra{\Psi_{(ab)}}\op{H}\Ket{\Psi_{(abc)}}.
\end{equation}
The constraints on the supersystem amplitudes are
\begin{equation}
t_{\mu\lambda} = \frac{1}{\sqrt{3}}
\label{eq:const1}
\end{equation}
and
\begin{equation}
t_{K\mu\lambda}  = \frac{-1}{\sqrt{2}}\left(
t_{K\mu} - t_{K\lambda}
\right)
\label{eq:const2}
\end{equation}
We emphasize that those amplitudes, $t_{Ka}$ and $t_{KLa}$, are equivalent to the amplitudes for the original system.
Interested readers are referred to ref. \citenum{Small2017} for the detailed derivation of this supersystem approach.

As shown in ref. \citenum{Small2009}, CCVB is capable of reaching the correct dissociation limit as long as UHF can. Its energy becomes exact in that limit as its energy is merely the sum of the restricted open-shell HF (ROHF) energy of each high-spin fragment. The strengths of CCVB are its size-consistency, spin-purity, and polynomial-scaling cost (which is dominated by integral transformation). One would expect CCVB to fail for systems where UHF fails to reach a proper dissociation limit and only generalized HF (GHF) can reach the correct asymptote among available single-determinant wavefunctions. We shall see such examples later in the paper and we will also show that the scope of CCVB for OS systems turns out to be much broader than that of UHF.
\subsection{Primer: The OS PP+i2 \revreplace{Ans{\"a}tz}{Ansatz}}
As mentioned in ref. \citenum{prep1}, a simple way to go beyond CCVB for closed-shell systems is to remove all the terms that contain amplitudes other than $t_{KL}$ when solving $R_{KL}=0$ in Eq. \eqref{eq:T2}. We refer \revreplace{this to}{to this} as the independent amplitude approximation (IAA) approach. This modified amplitude equation leads to a model called PP+i2. Unlike CCVB, PP+i2 is quite often variationally unstable (i.e., the resulting PP+i2 energy is too low). However, it can often reach correct asymptotes when CCVB cannot. Due to its simplicity, we tried to extend the PP+i2 \revreplace{ans{\"a}tz}{ansatz} to the open-shell systems and shall explain subtle difficulties involved in pursuing it below. However, we do not report any results associated with this model in this work.

The CS PP+i2 amplitude equation follows
\begin{equation}
^{(2)}R_{KL} 
\equiv
\langle \psi_{(KL)}| \hat{\mathcal H} | \xi_{(KL)}\rangle
- t_{KL}\langle \psi_0 | \hat{\mathcal H} | \xi_{(KL)}\rangle
\label{eq:csiaa2}
\end{equation}
where
\begin{equation}
| \xi_{(KL)}\rangle = | \psi_0\rangle + t_{KL}| \phi_{(KL)}\rangle
\end{equation}
Solving $^{(2)}R_{KL} = 0$ leads to a simple quadratic equation in $t_{KL}$ and different amplitudes are decoupled from each other. 
The solution to this quadratic equation might not exist and we observed this quite frequently near its variational breakdown. When the solution exists, we chose the one out of two solutions that gives a lower CCVB total energy. This approach was inspired by independent electron-pair approximations in coupled-cluster theory.\cite{Sinanoglu1962,Ahlrichs1968,Ahlrichs1968a,Ahlrichs1968b,Jungen1970,Ahlrichs1970,Gelus1971,Gelus1973,Lischka1973,Staemmler1973,Driessler1973,Dyczmons1974}

The natural inclination towards OS PP+i2 would be to solve (employing the supersystem approach) 
$^{(2)}\Omega_{Ka}  = 0$ where we define
\begin{align}
{}^{(2)}\Omega_{KL}  &= {}^{(2)}R_{KL} \\
{}^{(2)}\Omega_{K\mu} 
&=
\langle \Psi_{(K\mu)}| \hat{\mathcal H} | \Xi_{(K\mu)}\rangle
- t_{K\mu}\langle \Psi_0 | \hat{\mathcal H} | \Xi_{(K\mu)}\rangle
+ \sum_{\lambda\ne \mu} \kappa_{K\mu;\lambda} t_{K\mu\lambda}
\label{eq:osiaa1}
\end{align}
where
\begin{equation}
| \Xi_{(K\mu)}\rangle = | \Psi_0\rangle + t_{K\mu}| \Phi_{(K\mu)}\rangle
\end{equation}
One may impose the constraints in Eq. \eqref{eq:const1} and Eq. \eqref{eq:const2} so that the supersystem amplitudes represent the original system amplitudes. This is what we initially tried and it worked reasonably well.

Alternatively to this approach, one may attempt to apply the IAA approach to the original system amplitude equation and then apply the supersystem transformation. This is not as simple as what is described above as $\Ket{\psi_{(K\mu)}}$ and $\Ket{\psi_{(K\lambda)}}$ (or the corresponding vectors in the dual frame) are not necessarily orthogonal even when $\mu \ne \lambda$. With this in mind, we tried to allow $t_{K\mu}$ to depend on $t_{K\lambda}$ in the supersystem amplitude equation. This then leads to a modified CS-OS residual:
\begin{equation}
^{(2)}\tilde{\Omega}_{K\mu} 
=
\langle \Psi_{(K\mu)}| \hat{\mathcal H} | \tilde{\Xi}_{K}\rangle
- t_{K\mu}\langle \Psi_0 | \hat{\mathcal H} | \tilde{\Xi}_{K}\rangle
+ \sum_{\lambda\ne \mu} \kappa_{K\mu;\lambda} t_{K\mu\lambda}
\label{eq:osiaa2}
\end{equation}
where
\begin{equation}
| \tilde{\Xi}_{K}\rangle
=
|\Psi_0\rangle 
+ \sum_{\lambda} t_{K\lambda}| \Phi_{(K\lambda)}\rangle
\end{equation}
Solving $^{(2)}\tilde{\Omega}_{K\mu} =0$ under the constraints in Eq. \eqref{eq:const1} and Eq. \eqref{eq:const2} for $t_{K\mu}$ involves still a quadratic equation in $t_{K\mu}$ and one may employ an iterative approach to solve the residual equation until we have a self-consistent set of $\{t_{K\mu}\}$. This version of OS PP+i2 was found a little more variationally stable than the one described above, but it is still generally not recommended due to its instability. We use this OS PP+i2 to obtain a set of initial amplitudes for the subsequent CCVB calculations. 

In the case of CS fragments, size-consistency means that the energy of non-interacting {\it closed-shell} molecules is merely the sum of individual CCVB energy of each molecule. This is satisfied if every intermolecular amplitude is zero at the well-separated limit. Evidently, Eq. \eqref{eq:csiaa2} is size-consistent. In the case of OS fragments, we define the size-consistency as follows: the energy of non-interacting molecules (either {\it closed-shell} or {\it open-shell}) is the sum of their individual CCVB energies assuming that fragments are not spin-coupled into a lower spin manifold. Based on this definition, we conclude that both Eq. \eqref{eq:osiaa1} and Eq. \eqref{eq:osiaa2} ensure the size-consistency of PP+i2.

\subsection{The OS CCVB-3 \revreplace{Ans{\"a}tz}{Ansatz}}
Extending the CCVB wavefunction might seem relatively straightforward; we augment the cluster expansion of CCVB with missing 3P substitutions.
The first inclination might be to try the following CC expansion,
\begin{align}
\Ket{\psi_\text{3P}} &= \nonumber 
\Ket{\psi_\text{2P}}
+ \sum_{\substack{KLM\\K<L<M}}t_{KLM}\Ket{\phi_{(KLM)}}
+ \sum_{\substack{KL\mu\\K<L}}t_{KL\mu}\Ket{\phi_{(KL\mu)}}\\\nonumber
&+\sum_{\substack{KLPQR\\K<L<P<Q<R}} 
\left(t_{KL}t_{PQR} \Ket{\phi_{(KL)(PQR)}}
+ \{\text{nine other permutations}\}\right)\\\nonumber
&+\sum_{\substack{KLMN\mu\\K<L<M<N}} 
\left(t_{KL}t_{MN\mu} \Ket{\phi_{(KL)(MN\mu)}}
+ \{\text{five other permutations}\}\right)\\\nonumber
&+\sum_{\substack{KLMN\mu\\K<L<M<N}} 
\left(t_{K\mu}t_{LMN} \Ket{\phi_{(K\mu)(LMN)}}
+ \{\text{three other permutations}\}\right)\\\nonumber
&+\sum_{\substack{KLM\mu\lambda\\K<L<M\\\mu<\lambda}} 
(
t_{K\mu}t_{LM\lambda}
\Ket{\phi_{(K\mu)(LM\lambda)}}
+t_{L\mu}t_{KM\lambda}
\Ket{\phi_{(L\mu)(KM\lambda)}}
+t_{M\mu}t_{KL\lambda}
\Ket{\phi_{(M\mu)(KL\lambda)}}\\
&\:\:\:+ \{\text{3 terms from }(\mu\leftrightarrow\lambda)\}
)+\cdot\cdot\cdot.
\end{align}
We note that CCVB-3 includes all possible spin configurations through the cluster expansion. In other words, it is complete in the sense that the cluster expansion includes the same number of spin configurations as that of SCVB for a given active space. The remaining difference between SCVB and CCVB-3 is then largely from strong orthogonality between pairs assuming the CC approximation to the spin-coupling vector is reasonable.

Following the previous supersystem strategies, 
we first define the CCVB-3 energy as
\begin{equation}
E \equiv \Bra{\Psi_0}\hat{\mathcal H}\Ket{\Psi_\text{3P}}
\end{equation}
We then write the supersystem CS-CS 2P residual in the following manner:
\begin{equation}
^\text{3p}\Omega_{KL} = R_{KL} + \sum_{a\notin\{K,L\}}t_{KLa} \kappa_{KL;a}
\label{eq:iaa3t2cs}
\end{equation}
The CS-OS 2P residual can also be similarly defined:
\begin{equation}
^\text{3p}\Omega_{K\mu} = R_{K\mu} + \sum_{a\notin\{K,\mu\}}t_{K\mu a} \kappa_{K\mu;a}
\label{eq:iaa3t2os}
\end{equation}
Up to the 2P amplitude equation, the model is evidently size-consistent.

We further proceed to the 3P amplitude residual equation:
\begin{align}\nonumber
^\text{3p}\Omega_{KLa}
&=
 \Bra{\Psi_{(KLa)}}\hat{\mathcal H}\Ket{\Psi_\text{3P}} - Et_{KLa} \\\nonumber
&=
t_{KL}\Bra{\Psi_{(KLa)}}\hat{\mathcal H}\Ket{\Phi_{(KL)}}
+ t_{Ka}\Bra{\Psi_{(KLa)}}\hat{\mathcal H}\Ket{\Phi_{(Ka)}}
+ t_{La}\Bra{\Psi_{(KLa)}}\hat{\mathcal H}\Ket{\Phi_{(La)}}
\\\nonumber
&+\sum_{b\notin\{K,L,a\}} \left(
t_{KLb}\Bra{\Psi_{(KLa)}}\hat{\mathcal H}\Ket{\Phi_{(KLb)}}
+ t_{Kab}\Bra{\Psi_{(KLa)}}\hat{\mathcal H}\Ket{\Phi_{(Kab)}}
+ t_{Lab}\Bra{\Psi_{(KLa)}}\hat{\mathcal H}\Ket{\Phi_{(Lab)}}
\right)
\\\nonumber
&+\sum_{b\notin\{K,L,a\}}\left(
t_{Ka}t_{Lb}\langle \Psi_{(KLa)}|\hat{\mathcal H}|\Phi_{(Ka)(Lb)}\rangle
+t_{Kb}t_{La}\langle \Psi_{(KLa)}|\hat{\mathcal H}|\Phi_{(Kb)(La)}\rangle
+t_{KL}t_{ab}\langle \Psi_{(KLa)}|\hat{\mathcal H}|\Phi_{(KL)(ab)}\rangle
\right)
\\\nonumber
&+\sum_{b,c\notin\{K,L,a\}}\left(
t_{bc}t_{KLa}\langle \Psi_{(KLa)}|\hat{\mathcal H}|\Phi_{(bc)(KLa)}\rangle
+
\{\text{9 other permutations}\} 
\right)\\
&+t_{KLa}\left(\Bra{\Psi_{(KLa)}}\hat{\mathcal H}\Ket{\Phi_{(KLa)}}-E\right)
\label{eq:iaa3amp}
\end{align}
We now show that this residual equation may not yield a size-consistent model due to the contributions from the 5P substitutions.

We first assume that every amplitude that contains pairs associated with more than one fragment is all zero. We additionally assume that $K$ is well separated from $L$ and $a$ in Eq. \eqref{eq:iaa3amp}. Our goal is to check whether the residual equation in Eq. \eqref{eq:iaa3amp} is trivially zero under these assumptions. It is easy to see that the contributions from the 2P and 3P substitutions are zero either because the pertinent amplitude is zero or the hamiltonian matrix element is zero.
Moreover, the last term in Eq. \eqref{eq:iaa3amp} is zero as $t_{KLa} = 0$.

The 4P contribution is not as trivial to see that it is zero, so we discuss more details.
Among three terms listed in the summation of the 4P terms, only the second term may survive as $t_{Ka} = t_{KL} = 0$. We now claim that
\begin{equation}
t_{Kb}t_{La}\langle \Psi_{(KLa)}|\hat{\mathcal H}|\Phi_{(Kb)(La)}\rangle = 0
\label{eq:zeroquad}
\end{equation}
for any $b$. If $b$ is not on the same fragment as $K$, $t_{Kb}=0$ and thus the entire 4P terms are zero.
If $b$ is on the same fragment as $K$, the matrix element, $\langle \Psi_{(KLa)}|\hat{\mathcal H}|\Phi_{(Kb)(La)}\rangle$, is zero.
We provide more details on how to show that this matrix element is zero in the Appendix.

We turn our focus to the 5P contribution. We again assume that $K$ is well separated form $L$ and $a$. There are only three terms in the summation that are not trivially zero:
\begin{equation}
\sum_{b,c\notin\{K,L,a\}}\left(
t_{Kb}t_{Lac}\langle \Psi_{(KLa)}|\hat{\mathcal H}|\Phi_{(Kb)(Lac)}\rangle
+ 
t_{Kc}t_{Lab}\langle \Psi_{(KLa)}|\hat{\mathcal H}|\Phi_{(Kc)(Lab)}\rangle
+
t_{La}t_{Kbc}\langle \Psi_{(KLa)}|\hat{\mathcal H}|\Phi_{(La)(Kbc)}\rangle
\right)
\label{eq:i3consistency}
\end{equation}
The overlap matrix and its pseudoinverse that defines the transformation between $\Phi$-set and $\Psi$-set are given in Appendix.
First, we consider the case where $b$ and $c$ both are on a different fragment from that of $K$. In this case, all the associated amplitudes vanish so the corresponding contribution to Eq. \eqref{eq:i3consistency} is zero. 
Next, we consider the case where $b$ and $c$ both are on the same fragment as $K$. This automatically zeroes out the first two terms in Eq. \eqref{eq:i3consistency} as $t_{Lac} = t_{Lab}=0$. As the associated amplitudes are not zero for the third term, we must examine whether the integral vanishes. After some algebra, one can show that the third term does not vanish in general. 
Similarly, in the case where only one of $b$ and $c$ is on the same fragment as $K$ and the other one is on the same fragment as $L$ and $a$, the pertinent integral does not vanish in general. Therefore, we conclude that CCVB-3 is not size-consistent in general.

\subsection{The OS CCVB+i3 \revreplace{Ans{\"a}tz}{Ansatz}}

In the previous section, we proved that CCVB-3 is in general not size-consistent due to the 5P contributions in the amplitude equation. It is quite tempting to ignore those 5P contributions and build a model based on other terms up to the 4P contributions. Though this is an interesting model to try, an even simpler model can be formulated by applying the same strategy as that of OS PP+i2. Namely, $\Bra{\Psi_{(KLa)}}$ couples only with configurations containing exactly $(KLa)$ or its subset. We call this model along with the full treatment of 2P configurations (i.e., Eq. \eqref{eq:iaa3t2cs} and Eq. \eqref{eq:iaa3t2os}), CCVB+i3.

The supersystem OS CCVB+i3 amplitude residual then reads
\begin{align}
{}^{(3)}\Omega_{KLa}
&= \langle\Psi_{(KLa)}|\hat{\mathcal H}| \Xi_{(KLa)}\rangle - t_{KLa} \langle\Psi_{0}|\hat{\mathcal H}| \Xi_{(KLa)}\rangle
\label{eq:iepaT3}
\end{align}
where
\begin{equation}
| \Xi_{(KLa)}\rangle = 
| \Psi_0\rangle + t_{KL}| \Phi_{(KL)}\rangle
+ t_{Ka}| \Phi_{(Ka)}\rangle
+ t_{La}| \Phi_{(La)}\rangle
+ t_{KLa}| \Phi_{(KLa)}\rangle.
\label{eq:iaa1}
\end{equation}
Solving ${}^{(3)}\Omega_{KLa}=0$ for $t_{KLa}$ is quite straightforward as it is a simple linear equation in $t_{KLa}$. The solution follows
\begin{equation}
t_{KLa} = 
\frac{
t_{KL}\kappa_{KL;a}
-t_{Ka}\kappa_{Ka;L}
+t_{La}\kappa_{La;K}
}
{\left(
t_{KL} \mu_{KL}
+t_{Ka} \mu_{Ka}
+t_{La} \mu_{La}
\right)-\omega_{KLa}}
\label{eq:T3}
\end{equation}
where 
\begin{equation}
\omega_{KLa} = \Bra{\Psi_{(KLa)}}\hat{\mathcal H}\Ket{\Phi_{(KLa)}}-\Bra{\Psi_0}\hat{\mathcal H}\Ket{\Psi_0}
\end{equation}
and
\begin{equation}
\mu_{Ka}
=\Bra{\Psi_0}\hat{\mathcal H}\Ket{\Phi_{(Ka)}}.
\end{equation}
$\kappa_{Ka;b}$, $\mu_{Ka}$, and $\omega_{KLa}$ are expressed in terms of readily computable quantities given in Appendix.
We note that $t_{KLa} = -t_{KaL} = t_{LaK} = -t_{LKa} = t_{aKL} = -t_{aLK}$ in constrast to $t_{Ka} = t_{aK}$.

Similarly to OS PP+i2, there may be more than one way to choose $|\Xi_{KL\mu}\rangle$ due to the fact that configurations involving an OS electron are not orthogonal configurations involving different OS electrons or even of different substitution levels. For instance, 
we have 
$
\Braket{\psi_{(KL\mu)}|\psi_{(KL\lambda)}} \ne 0
$ when $\mu\ne\lambda$.
One may apply the IAA approach to the original system amplitude equation and then transform to the supersystem configurations.
In this case, the amplitudes associated with $\Ket{\psi_{(KL\mu)}}$ shall involve contributions from $\Ket{\psi_{(KL\lambda)}}$ as they overlap. 
This leads to a different choice of $|\Xi_{KL\mu}\rangle$,
\begin{equation}
| \tilde{\Xi}_{(KL\mu)}\rangle = | \Psi_0\rangle + t_{KL}| \Phi_{(KL)}\rangle
+ \sum_{\lambda}\left(
t_{K\lambda}| \Phi_{(K\lambda)}\rangle
+ t_{L\lambda}| \Phi_{(L\lambda)}\rangle
+ t_{KL\lambda}| \Phi_{(KL\lambda)}\rangle
\right)
\end{equation}
While this is certainly an interesting alternative, in this paper we focus on the CCVB+i3 model with Eq. \eqref{eq:iepaT3}.

In CCVB+i3, we work with only $t_{Ka}$ as independent variables and $t_{KLa}$ is directly parametrized by $t_{Ka}$. $t_{KLa}$ can be viewed as an attempt to incorporate the 3P influence using only 2P amplitudes. One may argue that $t_{KLa}$ should still be considered independent wavefunction parameters. Our viewpoint, however, is that $t_{KLa}$ in CCVB+i3 is not considered independent as it does not increase the computational scaling of CCVB. Hence, we claim that we have not effectively increased the number of independent wavefunction parameters going from CCVB to CCVB+i3. This contrasts with other wavefunction methods such as the second-order M{\o}ller-Plesset perturbation theory (MP2) and CC singles and doubles with a perturbative triples (CCSD(T)). In MP2, the doubles amplitudes are directly parametrized by HF orbitals. It, however, has a steeper computational scaling compared to HF. Similarly, in CCSD(T), the perturbative triples amplitudes are directly parametrized by singles and doubles amplitudes with an increase in the computational scaling.

The scope of CCVB+i3 is not yet clear to us even though we have not yet found a system where CCVB+i3 fails to dissociate properly. Mathematical proofs related to this and a more complicated and improved CCVB model will be investigated in the future.

\subsection{CCVB+i3 Lagrangian and Orbital Optimization}
We establish the Lagrangian of CCVB+i3 for the orbital optimization. The Lagrangian follows
\begin{equation}
\mathcal L
=
E
+
\sum_{K<a}
\lambda_{Ka}
\left(
R_{Ka}
+
\sum_{b\notin \{K,a\}} 
t_{Kab} \kappa_{Ka;b}
\right)
+
\sum_{K<L<a}
\lambda_{KLa}
{}^{(3)}\Omega_{KLa}
\label{eq:lag}
\end{equation}
where $\lambda_{Ka}$ and $\lambda_{KLa}$ represent the $L$-amplitudes.
As we know $t_{KLa}$ as a function of $t_{KL}, t_{Ka},$ and $t_{La}$, one may avoid using $\lambda_{KLa}$ and directly substitute the result to $t_{Kab}$ above. However, we chose to work with $\lambda_{KLa}$ as a matter of convenience.
In Eq. \eqref{eq:lag}, we left out $\lambda_{\mu\nu}$ and $\lambda_{K\mu\nu}$ as their $T$-amplitudes are constrained;
we incorporate $t_{\mu\nu}$ and $t_{K\mu\nu}$ through explicit substitutions to the Lagrangian using Eq. \eqref{eq:const1} and Eq. \eqref{eq:const2}.
Evidently, we have
\begin{equation}
\frac{\partial \mathcal L}{\partial \lambda_{Ka}} = 0 = \Omega_{Ka}
\end{equation}
and
\begin{equation}
\frac{\partial \mathcal L}{\partial \lambda_{KLa}} = 0 = {}^{(3)}\Omega_{KLa}
\end{equation}
From
\begin{equation}
\frac{\partial \mathcal L}{\partial t_{Ka}} = 0,
\end{equation}
we obtain
\begin{equation}
0 = 
\frac{\partial E}{\partial{t_{Ka}}}
+
\sum_{M<b}
\lambda_{Mb}
\frac{\partial \Omega_{Mb}}
{\partial t_{Ka}}
+
\sum_{b\notin\{K,a\}}
\lambda_{Kab}
\frac{\partial {}^{(3)}\Omega_{Kab}}
{\partial t_{Ka}}
\label{eq:L2}
\end{equation}
where in the last sum we have $\lambda_{K\mu\lambda} = 0$ as discussed before.
Those derivatives in Eq. \eqref{eq:L2} are provided in terms of computable quantities in the Appendix. 
We obtain $\lambda_{KLa}$ from
\begin{equation}
\frac{\partial \mathcal L}{\partial t_{KLa}} = 0,
\end{equation}
which yields
\begin{equation}
\lambda_{KLa}
=
\frac{
\lambda_{KL}\kappa_{KL;a}
-\lambda_{Ka}\kappa_{Ka;L}
+\lambda_{La}\kappa_{La;K}
}
{
t_{KL}\mu_{KL}
+
t_{Ka}\mu_{Ka}
+
t_{La}\mu_{La}
-\omega_{KLa}
}
\label{eq:L3}
\end{equation}
We note that $\lambda_{KLa}$ is antisymmetric under permuting two indices as in $t_{KLa}$. We substitute $\lambda_{KLa}$ in Eq. \eqref{eq:L2} using Eq. \eqref{eq:L3}. The resulting equation is only linear in $\lambda_{Ka}$ and therefore $\lambda_{Ka}$ can be uniquely determined as long as Eq. \eqref{eq:L2} is not ill-defined.

Having solved the $L$-amplitude and $T$-amplitude equations, the subsequent orbital optimization is relatively straightforward. We parametrize orbital rotations with a unitary exponential matrix,
\begin{equation}
\mathbf C = \mathbf C_0 \exp\left({\Delta - \Delta^\dagger}\right)
\end{equation}
and the pertinent orbital gradient and Hessian are obtained taking derivatives of the Lagrangian in Eq. \eqref{eq:lag}.
The orbital gradient reads
\begin{equation}
L_{pq}^{\Delta} = \frac {\partial {\mathcal L}} {\partial{\Delta_{pq}}}
\label{eq:grad}
\end{equation}
and the Hessian reads
\begin{equation}
H_{pq,rs}^{\Delta\Delta} = \frac {\partial {\mathcal L}} {\partial{\Delta_{pq}}\partial{\Delta_{rs}}}
\label{eq:hess}
\end{equation}
We also need the gradient and the hessian of $\mathcal L$ with respect to the polarization angle $\theta_K$ (i.e., $\vec L^\theta$ and $\mat H^{\theta\theta}$) and those can be obtained in the exactly same fashion. We treat $\Delta_{pq}$ and $\theta_K$ as independent variables and optimize the Lagrangian over those parameters. 
These are enough to establish any first-order convergence techniques and the optimizer we employed in this work is the geometry direct minimization (GDM) \cite{Voorhis2002,Dunitez2002,Lawler2010} which needs an orbital gradient and the diagonal elements of Hessian.
Most of the relevant terms in $\mat L^\Delta$, $\mat H^{\Delta\Delta}$, $\vec L^\theta$, and $\mat H^{\theta\theta}$ are available in ref. \citenum{Small2014}. We discuss the CCVB+i3 specific terms in Appendix.
\subsection{Amplitude Solvers}
We found solving the CCVB+i3 amplitude equations quite challenging for some systems presented below. 
This may be understood by observing that there are many low-lying states nearly degenerate for strongly correlated systems and this
implies that there is more than one set of cluster amplitudes that can represent the state of our interest.
Unlike linear wavefunctions, this poses a great challenge to CC wavefunctions as amplitude solvers may get easily lost during iterations and amplitude equations may become nearly ill-conditioned. Therefore, we had to try several different solvers discussed below.

For CCVB calculations, the recommended $T$-amplitude and $L$-amplitude solver is Gauss-Seidel combined with Pulay's direct inversion of the iterative space (GS-DIIS).\cite{Pulay1980,Pulay1982} The GS step solves a quadratic equation in Eq. \eqref{eq:T2amp1} and Eq. \eqref{eq:T2amp2} for $t_{Ka}$ while keeping all other amplitudes fixed. It solves a linear equation in the case of $\lambda_{Ka}$. The GS step scales cubically with the number of pairs. One may consider using GS-DIIS for CCVB+i3 when solving Eq. \eqref{eq:iaa3t2cs} and Eq. \eqref{eq:iaa3t2os}. It becomes a cubic equation in $t_{Ka}$ once one multiplies both sides by the denominator of Eq. \eqref{eq:T3}. It is still a linear equation in the case of $\lambda_{Ka}$. The computational cost scales still cubically with the system size.

When GS-DIIS fails to find a solution, we employ the Gauss-Newton method with line-search (GN-LS). This was done by defining a cost function $f$ as a squared sum of amplitude equations. In other words, the cost function reads
\begin{equation}
f = \frac12 \sum_{K<a} |\Omega_{Ka}|^2
\label{eq:squaresum}
\end{equation}
The line-search guarantees a descent direction that decreases the value of $f$. The GN-LS method requires the evaluation of Jacobian $\mat J$ and the inverse of it. The Jacobian of CCVB and CCVB+i3 are given in Appendix. As the length of $\mat J$ scales quadratically with the number of pairs, $n_p$, it requires $\mathcal O(n_p^4)$ amount of work to evaluate it and $\mathcal O(n_p^6)$ to invert it. While this is not an ideal solver for CCVB due to the steep scaling, for the systems studied in this paper, the time for solving amplitudes with GN-LS is negligible compared to that of integral transformation. 

Some systems exhibited serious numerical issues with GN-LS because $\mat J$ was nearly singular (i.e., the smallest singular value is roughly 1e$^{-6}$--1e$^{-7}$). In such cases, we found it more effective to use preconditioned limited-memory Broydon-Fletcher-Goldfarb-Shanno (L-BFGS) with line-search (L-BFGS-LS) where we used the inverse of the diagonal elements of $\mat J \mat J^T$ as the preconditioner. The evaluation of $\mat J$ is the bottleneck in this case which scales $\mathcal O(n_p^4)$. With this solver, the overall cost of CCVB methods is still dominated by integral transformation.

In this work, we used L-BFGS-LS for the numbers reported and GN-LS for testing purposes. L-BFGS-LS has been adequate for most systems described here, but when $\mat J$ is nearly singular its convergence becomes extremely slow. In fact, there is no bulletproof method when $\mat J$ is nearly singular; this is an interesting open question in applied mathematics.\cite{Nocedal2006}
\subsection{Computational Cost}
The computational cost of CCVB is dominated by integral transformation and the rest of the computations scale cubically with the system size if the amplitude equation is solved via GS-DIIS. \cite{Small2009} The cost of evaluating $t_{KLa}$ in Eq. \eqref{eq:T3} is dominated by the computation of $\omega_{KLa}$. A naive way to evaluate $\omega_{KLa}$ would scale quartically due to the summation involved in Eq. (S18).
However, if we precompute $\sum_{a} \left(\sigma_{Ka;t_1t_1;ss} - \sigma_{Ka;ss;ss}\right)$ (quantities defined in Appendix) and store this for every CS pair $K$, we can evaluate $\omega_{KLa}$ with a cubic amount of work. Thus, CCVB+i3, in principle, scales the same as CCVB as long as the underlying amplitude solver takes an equal or less amount of work.
\section{Computational Details}
All CCVB, CCVB+i3, GVB-PP, spin-flip complete active space (SF-CAS),\cite{Mayhall2014,Mayhall2014a} HF, and self-consistent field molecular interaction (SCF-MI) \cite{Stoll1980,Khaliullin2006} calculations were performed with the development version of $\texttt{Q-Chem}$.\cite{Shao2015} CASSCF calculations were performed using $\texttt{Orca}$ \cite{Neese2017} and $\texttt{PySCF}$. \cite{Sun2017} Heat-bath CI (HCI) \cite{Holmes2016} and HCISCF \cite{Smith2017} calculations were carried out with $\texttt{Dice}$\cite{Dice} interfaced with $\texttt{PySCF}$. We also used $\texttt{GAMESS}$,\cite{Schmidt1993} and $\texttt{Psi4}$ \cite{Parrish2017} to crosscheck some of the CASSCF numbers presented below.

For all CCVB calculations but those for the [\ce{Cr9}] molecular magnet, we used a tolerance of $10^{-12}$ for the amplitudes solver, which tests the root-mean-square (RMS) of $\mat J \Omega$, and a tolerance of $1e^{-5}$ for the orbital optimizer, which tests the RMS of orbital gradients and the step size. These were enough to get energy converged up to 0.1-1 $\mu \text{H}$.
Due to numerical challenges we faced in studying [\ce{Cr9}], we used looser convergence criteria for [\ce{Cr9}]; $10^{-8}$-$10^{-10}$ for $T$-amplitudes. These were enough to converge energies up to 0.1-0.01 kcal/mol.
The CASSCF calculations were converged up to at least 0.1 $\mu \text{H}$. 
All plots were generated with $\texttt{Matplotlib}$. \cite{Hunter2007} All molecular figures were generated with $\texttt{Chemcraft}$.\cite{Chemcraft}

We mention how we obtain an initial set of orbitals to perform CCVB calculations. Like most other pairing methods, the CCVB energy is not orbital-invariant and CCVB exhibits multiple local minima in orbital optimization.
Therefore, obtaining physically correct initial orbitals and pairing them properly in the beginning are often crucial to properly run CCVB calculations.
The procedure of obtaining an initial guess used in this work and running ``spin-ladder'' calculations is as follows:
\begin{enumerate}
\item We perform an SCF-MI calculation of the lowest spin state that UHF can describe for a given active space. 
For the examples discussed below, this state is always $M_S=3/2$.
\item We then perform a GVB-PP calculation of $S=3/2$ using the orbitals from the SCF-MI calculation.
Orbitals from this GVB-PP calculation are well localized.
\item We use chemical intuition, usually based on sensible Lewis dot structures, to pair localized orbitals properly for $S=1/2$. This is the most non-trivial step when running CCVB. 
\item We run GVB-PP to optimize the new pairing for the lowest spin-state, in our case $S=1/2$.
\item We run CCVB for $S=1/2$ using GVB-PP orbitals from Step 4.
\item From the solution of $S=1/2$, we unpair most polarized pairs based on their polarization angles to obtain initial orbitals for higher spin states. We found this ``spin-ladder'' calculation quite robust. This approach was also used in ref. \citenum{Small2017}.
\end{enumerate}
For those systems discussed below, this procedure always produced a sensible solution. Due to the ambiguity involved in Step 3, one may try multiple possible Lewis dot structures in general. 
\section{Results and Discussions}
Spin-frustrated systems often exhibit a UHF to GHF instability, so one may expect CCVB to fail qualitatively for those. In the following, we consider a total of five spin-frustrated systems. Those systems are spin-frustrated based on the Kahn's definition\cite{Kahn1997}; they are open-shell, involve an odd number of sites, and are geometrically symmetric. We consider systems with three unpaired electrons per site which correspond to the S-$\frac32$ Heisenberg model. This choice was made based on observations from S-$\frac12$ frustrated systems such as a hydrogen lattice where UHF can dissociate the lowest spin-state correctly. Consistent with the success of UHF, we did not observe any significant differences between CCVB and CCVB+i3 in this case.

Studying S-$\frac32$ systems, we uncover that 3P configurations are necessary for a qualitatively correct description of the lowest spin state (i.e., doublet) and even higher spin states in some cases.
We emphasize the role of $\epsilon_{s5}$ and $\epsilon_{d5}$, which are missing in CCVB. In terms of the original system configurations, $\epsilon_{d5}$ is generated by annihilating two singlet pairs and an $\alpha$ electron and creating two triplet pairs coupled to a triplet which is then recoupled with the remaining unpaired electron to form an overall doublet. This configuration is captured by the corresponding supersystem configuration. The detailed discussion on the role of the $\epsilon_{s5}$ configurations in CS systems is given by two of us.\cite{prep1}

\subsection{\ce{N3} ($\text{D}_\text{3h}$) -- Spin-frustration involving p orbitals}
Triangular \ce{N3} is perhaps the simplest system that satisfies our criteria described above. It has three unpaired electrons per site and is strictly spin-frustrated due to symmetry. We consider its dissociation to three nitrogens ($^4$S) within the cc-pVTZ basis set.\cite{Dunning1989} The active space we consider is (9e, 9o) and it is small enough to perform exact CASSCF calculations. For the $S=9/2$ state, ROHF is exact within the active space and thus CCVB is also exact. There are no $|\Phi_{(KLa)}\rangle$ configurations in the $S=7/2$ state. Hence, CCVB and CCVB+i3 are exactly identical in the case of $S=7/2$. UHF can properly dissociate $M_S=3/2$ and $M_S=9/2$.
\begin{figure}[h!]
\includegraphics[scale=0.55]{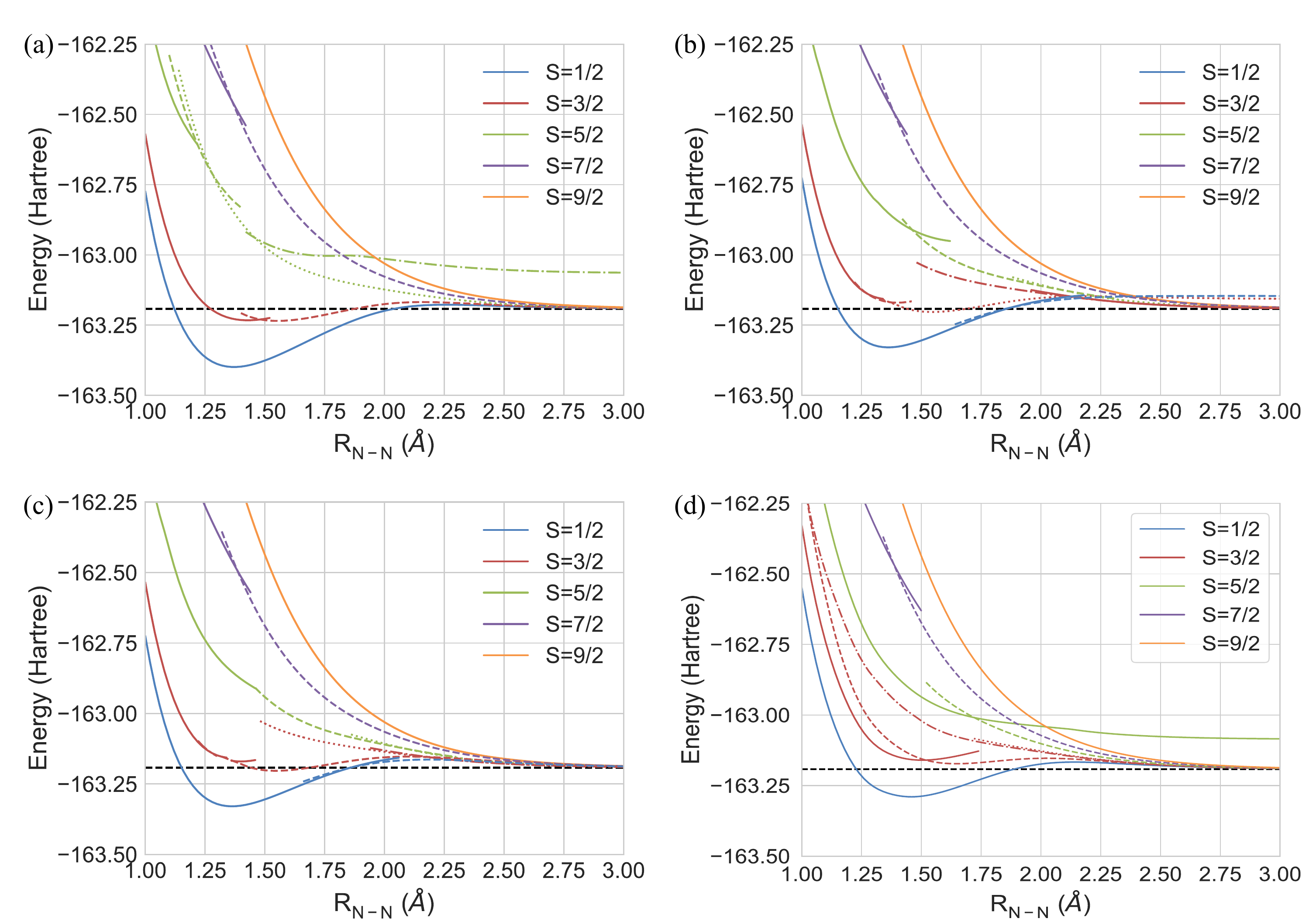}
\caption{\label{fig:n3pes}
Potential energy curves of \ce{N3} from
  {(a)} CASSCF (9e, 9o),
  {(b)} CCVB,
  {(c)} CCVB+i3, \revinsert{and {(d)} SF-CAS.}
The black dotted line indicates the asymptotic energy of three \ce{N}($^4$S), $E = -163.1920735\: E_h$.
Different line styles for each spin state represent different local minima.
}
\end{figure}

We first discuss potential energy curves (PECs) of CASSCF(9e,9o) calculations for each spin state as illustrated in Figure \ref{fig:n3pes}. Only $S=1/2$ and $S=3/2$ are bound states with
a small hump towards the dissociation limit.
The $S=3/2$, $S=5/2$, and $S=7/2$ states exhibit multiple solutions, which may indicate an inadequate choice of active space.
Moreover, those CASSCF solutions break the spatial symmetry. 
One could employ a larger active space to see whether it resolves these issues, but we focus on only pairing active spaces of the form of ($n$e, $n$o) for the purpose of this work. We note that we observed multiple solutions even from CASSCF(15e, 12o) calculations, which are full valence active space calculations. This indicates that those solutions may actually be physical and there may exist state crossings. However, this larger active space CASSCF still breaks spatial symmetry except for $S=9/2$ (ROHF limit). \revreplace{It would be interesting to investigate those further in the future to see whether the full valence active space is actually suitable for this problem.}{}

Two CASSCF solutions in $S=3/2$ show small differences in converged orbitals. We employed Knizia's intrinsic bond orbital (IBO) localization scheme \cite{Knizia2013} to characterize two solutions at $R=1.4$ \AA. As we localized only the active space orbitals, this procedure does not change the CASSCF energy. The solid line solution has one bond-like orbital whereas the dotted line solution shows only localized orbitals. This is quite sensible given that the dotted line solution is connected to the dissociation limit where localized orbitals are most sensible. There is a third solution that appears between 2.42 \AA{} and 3.00 \AA. This solution is almost identical to the dotted line and the energy difference between those two is less than 1 mE$_\text{h}$. We did not include this solution for simplicity.

For $S=5/2$, there are a total of five CASSCF solutions found. Interestingly, one of them does not dissociate properly. This solution involves a delocalized orbital even after the IBO localization. Its natural orbital occupation number indicates that there is a doubly occupied orbital and an empty orbital in the active space at $R=3.0$ {}\AA. One may suggest that this solution is dissociating to one \ce{N} ($^2$D) and two \ce{N} ($^4$S)'s, but its energy is about 16 kcal/mol higher than this limit at $R=3.0$ \AA. We suspect that it is an unphysical solution that comes from the delocalized orbital. 

Lastly, there are two solutions observed in the $S=7/2$ state. We compared orbitals of two solutions at $R=1.2$ \AA{} and the IBO localization analysis reveals more localized character in the higher energy solution (dotted) than in the lower energy solution (solid). Also, there is an almost doubly occupied orbital in the solid line based on natural orbital occupation numbers whereas the dotted line exhibits no such strong double occupation. It is sensible that the dashed line solutions \revinsert{are} indeed lower in energy when approaching the dissociation limit.

Both CCVB and CCVB+i3 in Figure \ref{fig:n3pes} successfully capture qualitative features of CASSCF solutions. 
Perhaps, the most interesting finding of two panels, (b) and (c), is that CCVB+i3 reaches the exact dissociation limit for every spin state while CCVB cannot reach the correct asymptote for $S=1/2$ and has some solutions for $S=3/2$ that cannot dissociate properly. This observation will be elaborated in greater detail \revreplace{below}{later}.

\revinsert{Lastly, we present SF-CAS results in Figure \ref{fig:n3pes} (d). Since it is based on spatially symmetric high-spin $S=9/2$ orbitals, these CAS wavefunctions are spatially symmetric. The low-spin solutions are lacking in orbital relaxation so these results are upper-bounds for symmetry-adapted CASSCF solutions.
Other than $S=1/2$, there are numerous solutions crossing and these are very similar to broken symmetry CASSCF, CCVB, and CCVB+i3 solutions.
}

Both CCVB methods involve two solutions in the $S=1/2$ state. We inspected the orbitals from two solutions at $R=2.0$ \AA. There are a pair of orbitals and a singly occupied orbital that are of very different character in each solution. The solid line, which is higher in energy at this geometry, has more delocalized orbitals while the dotted line exhibit more localized orbitals. 
There are more solutions than CASSCF in the case of $S=3/2$ and two of those solutions resemble those of CASSCF. The rest of solutions exhibit a purely repulsive curve which are likely unphysical. Reading those repulsive solutions into CASSCF, we confirmed that they are not close to any stable stationary points and they all collapse to the other solutions we have. The solutions for $S=5/2$ and $S=7/2$ can be easily compared to their CASSCF counterparts.
In passing, we note that those that appear in CCVB but not in CASSCF can be \insertnew{tentatively} attributed to \insertnew{the pairwise nonorthogonality limitation of CCVB rather than the spin-coupling limitation}. Therefore they are likely to disappear if we use its orbital-invariant generalization, CCVB-SD.\cite{Small2012,Lee2017}

\begin{figure}[h!]
\includegraphics[scale=0.52]{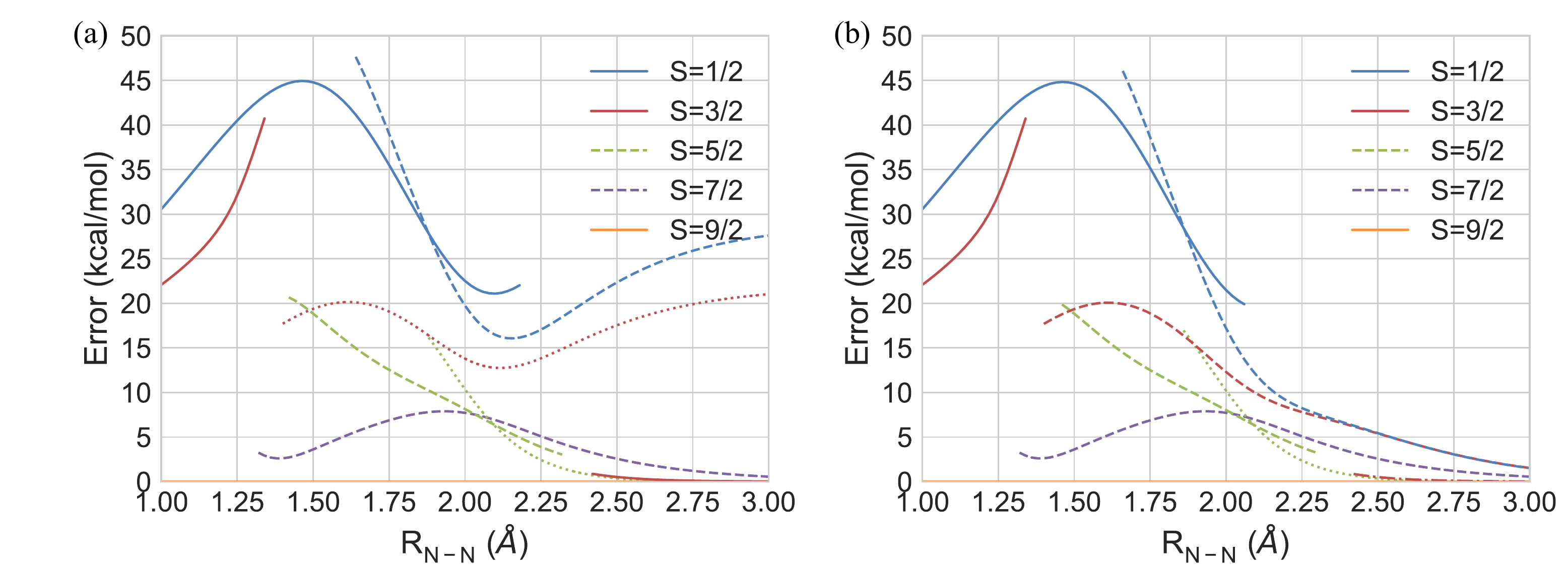}
\caption{\label{fig:n3error}
Errors in absolute energy relative to CASSCF of \ce{N3} for
  {(a)} CCVB and
  {(b)} CCVB+i3.
The line style of each line is consistent with that of Figure \ref{fig:n3pes}, and we omitted solutions that we could not find the counterpart in CASSCF.
}
\end{figure}
We present a more precise error analysis of two CCVB models against CASSCF results in Figure \ref{fig:n3error}. Near the equilibrium distance of $S=1/2$, all the states exhibit quite substantial \insertnew{CCVB} errors and this is a manifestation of the lack of ionic configurations relevant to dynamic correlations. However, as mentioned earlier, it should be emphasized that CCVB+i3 can dissociate all the spin states exactly in this example. CCVB shows two distinct solutions for $S=1/2$ and $S=3/2$ that do not dissociate properly. This is indeed the hallmark of 3P substitutions that are necessary to describe the spin frustration.

We also note that there are $S=5/2$ solutions in both CCVB and CCVB+i3, which exhibit a slight non-variationality (about $0.05$ kcal/mol) at $R=3.0$ \AA. This is the first time for us to observe non-variationality of CCVB, and we further confirmed this by reading CCVB orbitals into a CASSCF calculation and observing higher energy in the final CASSCF energy. 
As the extent to which CCVB manifests this non-variationality is almost negligible, we did not find it very concerning.
\begin{figure}[h!]
\includegraphics[scale=0.35]{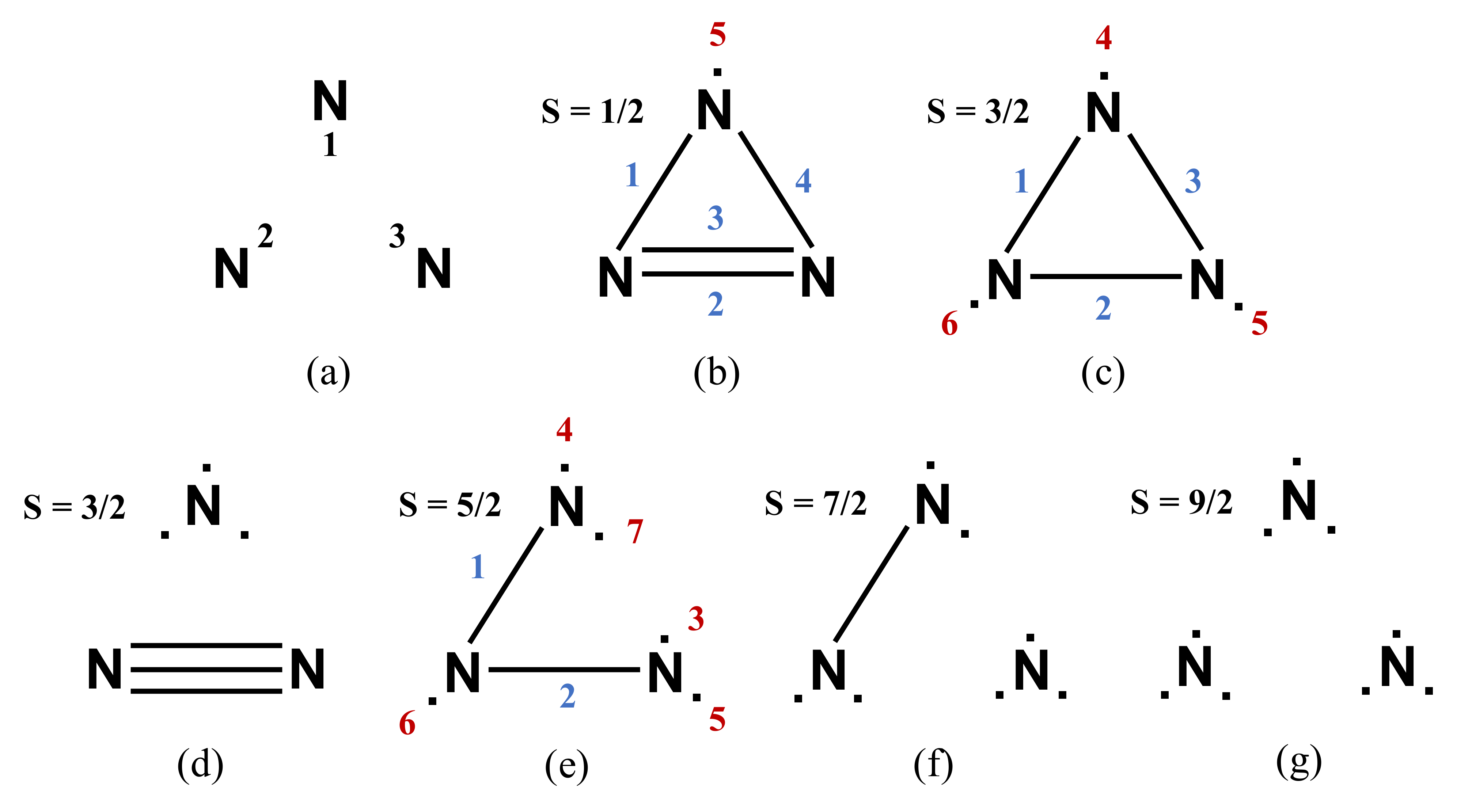}
\caption{\label{fig:n3lewis}
Panel (a) shows how we label the three nitrogens.
The \replacewith{rest}{remaining} panels illustrate possible Lewis structures of \ce{N3} for each spin state:
(b) $S=1/2$, (c) $S=3/2$, (d) $S=3/2$, (e) $S=5/2$, (f) $S=7/2$, and (g) $S=9/2$.
The number next to each bond or unpaired electron is used to label a CS (blue) or OS (red) pair.
(c) and (d) represent two possible Lewis structures of \ce{N3} when $S=3/2$.
Every N-N bond consists of p-orbitals and each N atom has three p-orbitals which yields an active space of (9e, 9o).
}
\end{figure}

It is interesting to discuss what the significant 3P substitutions are when dissociating this molecule.
As CCVB+i3 does not make significant differences near equilibrium bond lengths, CCVB and CCVB+i3 energies are very similar. 
However, $t_{KLa}$ becomes quite significant as one stretches bonds.

In Table \ref{tab:n3s2}, we \replacewith{discuss}{present} the 2P and 3P amplitudes of the $S=1/2$ CCVB+i3 solution at $R=3.0$ \AA{}. Orbitals are strongly localized, so each CS or OS pair corresponds to bonds and an unpaired electron in the Lewis structures in Figure \ref{fig:n3lewis}. Most of the $t_{Ka}$ amplitudes have values close to $\pm1/\sqrt{3}$, but there are two amplitudes that are nearly zero. Those two correspond to the $t_{K\mu}$ type amplitudes, where the CS pair $K=2,3$ is not connected 
to the radical center, N(1) (See Figure \ref{fig:n3lewis} (a) and (b)). 
$t_{K\mu}$ involving a CS pair connecting N(1) and others all exhibit quite large values. All the $t_{KL}$ amplitudes are large.

The two largest 3P amplitudes are of the $t_{KLM}$ type. Those CS pairs form a perfect triangle and this is typical of large $t_{KLM}$. Those involving three CS pairs that do not form a triangle are almost negligible as shown in Table \ref{tab:n3s2}. There are a total of four significant $t_{KL\mu}$ amplitudes. The basic observation is that they all involve one CS pair that connects N(1) with N(2) or N(3) and the second CS pair should connect N(2) and N(3). All the other $t_{KL\mu}$ amplitudes that do not satisfy this condition are all negligible.

\begin{table}
  \centering
\begin{tabular}{rr|r}\hline
$K$&\multicolumn{1}{c}{$a$}\vline&\multicolumn{1}{c}{$t_{Ka}$}\\
\hline
2    &     3    &    0.56500\\
4    &     5    &   -0.55826\\
1    &     5    &   -0.55825\\
3    &     4    &   -0.55449\\
1    &     3    &    0.55448\\
2    &     4    &   -0.53524\\
1    &     2    &    0.53523\\
1    &     4    &    0.53192\\
3    &     5    &   -0.00028\\
2    &     5    &   -0.00026\\
\hline
\end{tabular}
  \quad
  \begin{tabular}{rrr|r}\hline
$K$ & \multicolumn{1}{c}{$L$} & \multicolumn{1}{c}{$a$}\vline & \multicolumn{1}{c}{$t_{KLa}$} \\
\hline
1     &    3    &     4     &   0.74396\\
1     &    2    &     4     &   0.71019\\
3     &    4    &     5     &  -0.38253\\
1     &    3    &     5     &  -0.38213\\
2     &    4    &     5     &  -0.36031\\
1     &    2    &     5     &  -0.35995\\
2     &    3    &     4     &   0.00260\\
1     &    2    &     3     &   0.00260\\
2     &    3    &     5     &   0.00015\\
1     &    4    &     5     &  0.00000\\
\hline
  \end{tabular}
  \caption{The CCVB+i3 2-pair and 3-pair amplitudes of \ce{N3} ($S=1/2$) at $R=3.0$ \AA.
  1--4 are CS pairs and 5 is an OS pair. Pair labels are consistent with those in Figure \ref{fig:n3lewis} (b).
  }
  \label{tab:n3s2}
\end{table}

The $S=3/2$ state involves two reasonable Lewis dot structures, Figure \ref{fig:n3lewis} (c) and (d). Using the PP references that represent those Lewis dot structures yields different CCVB solutions. The (c) orbitals give a reasonable description near equilibrium, but CCVB cannot dissociate this solution to the right limit \insertnew{(neither can UHF)}. (c) involves a triangle and three pair substitutions become crucial to correctly dissociate. 
There is no triangle involved in the bonding network of (d), so the 3P amplitudes are expected to be negligible and CCVB \insertnew{(and UHF)} can properly dissociate. Therefore, CCVB can dissociate $S=3/2$ exactly with orbitals from (d), but it gives a purely repulsive potential energy. 
Table \ref{tab:n3s4} presents $t_{Ka}$ and $t_{KLa}$ for CCVB+i3 calculation using the (c) orbitals. The observation here is consistent with what is discussed above in the case of $S=1/2$.

\begin{table}
  \centering
\begin{tabular}{rr|r}\hline
$K$&\multicolumn{1}{c}{$a$}\vline&\multicolumn{1}{c}{$t_{Ka}$}\\
\hline
     1    &     6    &    0.55779\\
     1    &     4    &   -0.55779\\
     3    &     4    &    0.55779\\
     2    &     6    &   -0.55779\\
     2    &     5    &    0.55779\\
     3    &     5    &   -0.55779\\
     1    &     3    &   -0.53284\\
     1    &     2    &   -0.53284\\
     2    &     3    &   -0.53284\\
     2    &     4    &   -0.00001\\
     3    &     6    &    0.00001\\
     1    &     5    &    0.00000\\
\hline
\end{tabular}
  \quad
  \begin{tabular}{rrr|r}\hline
$K$ & \multicolumn{1}{c}{$L$} & \multicolumn{1}{c}{$a$}\vline & \multicolumn{1}{c}{$t_{KLa}$} \\
\hline
     1     &    2     &    3     &   0.70949\\
     1     &    2     &    4     &   0.36038\\
     1     &    3     &    6     &   0.36038\\
     1     &    2     &    5     &  -0.36038\\
     1     &    3     &    5     &  -0.36038\\
     2     &    3     &    6     &   0.36037\\
     2     &    3     &    4     &  -0.36037\\
     1     &    3     &    4     &  -0.00000\\
     2     &    3     &    5     &   0.00000\\
     1     &    2     &    6     &   0.00000\\
\hline
  \end{tabular}
  \caption{The CCVB+i3 2-pair and 3-pair amplitudes of \ce{N3} ($S=3/2$) at $R=3.0$ \AA.
  The solution here is well represented by the Lewis structure in Figure \ref{fig:n3lewis} (c).
  1--3 are CS pairs and 4--6 are OS pairs. Pair labels are consistent with those in Figure \ref{fig:n3lewis} (c).
  }
  \label{tab:n3s4}
\end{table}

The $S=5/2$ state does not exhibit any notable 3P contributions as shown in Table \ref{tab:n3s6}. This is particularly interesting because it contains $t_{125}$ and $t_{124}$ that are significant in describing the dissociation of $S=3/2$. The initial guess orbitals are from $S=1/2$ orbitals and we unpair two most polarized electron pairs to obtain the Lewis structure in Figure \ref{fig:n3lewis} (e). 
The effect of orbital optimization is very small at this distance.
However, once those localized initial orbitals are optimized, they become delocalized. The converged orbitals show almost no 3P contributions. 
Surprisingly, even the localized initial orbitals do not exhibit significant 3P contributions.
Since the energy difference between those two orbitals is only 1 mE$_\text{h}$, the amplitudes in Table \ref{tab:n3s6} are evaluated with those localized guess orbitals as a matter of convenience. 
\begin{table}
  \centering
\begin{tabular}{rr|r}\hline
$K$&\multicolumn{1}{c}{$a$}\vline&\multicolumn{1}{c}{$t_{Ka}$}\\
\hline
2    &     5    &   -0.56231\\
2    &     3    &   -0.56201\\
1    &     7    &   -0.56021\\
1    &     4    &   -0.55980\\
2    &     6    &    0.55930\\
1    &     6    &    0.55754\\
1    &     2    &    0.55043\\
2    &     7    &   -0.54007\\
2    &     4    &   -0.53817\\
1    &     3    &   -0.52954\\
1    &     5    &   -0.52867\\
\hline
\end{tabular}
  \quad
  \begin{tabular}{rrr|r}\hline
$K$ & \multicolumn{1}{c}{$L$} & \multicolumn{1}{c}{$a$}\vline & \multicolumn{1}{c}{$t_{KLa}$} \\
\hline
1     &    2     &    5    &    0.01945\\
1     &    2     &    3    &    0.01861\\
1     &    2     &    4    &   -0.01249\\
1     &    2     &    7    &   -0.01122\\
1     &    2     &    6    &   -0.00010\\
\hline
  \end{tabular}
  \caption{The CCVB+i3 2-pair and 3-pair amplitudes of \ce{N3} ($S=5/2$) at $R=3.0$ \AA.
  The orbitals used here are well represented by the Lewis structure in Figure \ref{fig:n3lewis} (e).
  1 and 2 are CS pairs and 3--7 are OS pairs. Pair labels are consistent with those in Figure \ref{fig:n3lewis} (e).
  }
  \label{tab:n3s6}
\end{table}

Furthermore, the amplitudes presented in Table \ref{tab:n3s6} show \insertnew{differences relative} to our previous observations from $S=1/2$ and $S=3/2$.
All the $t_{Ka}$ amplitudes are close to $\pm1/\sqrt{3}$, which include amplitudes involving an OS pair centered on N(1) and a CS pair connecting the other two nitrogens. Moreover, this change in $t_{Ka}$ essentially nullifies every $t_{KLa}$. For instance, we have non-negligible $t_{125}$ in $S=3/2$, but it is very small in $S=5/2$.
We compared every parameter involved in evaluating $t_{125}$ through Eq. \eqref{eq:T3}, and the only significant difference is that $t_{15}$ is zero in $S=3/2$, but is large in $S=5/2$. The same applies to $t_{124}$.
We believe that those large $t_{K\mu}$ amplitudes may be relevant to the broader applicability of CCVB than that of UHF for OS systems, but we do not have a clear way to understand the limit of its applicability yet. \insertnew{By contrast, it should be clear that UHF cannot dissociate $S=5/2$.}

\subsection{\ce{V3O3} ($\text{D}_\text{3h}$) -- Spin-frustration involving s and d orbitals}
Vanadium oxides have drawn a lot of attention from the solid state physics community and they are often strongly correlated. In particular, \ce{VO2} has been used to study metal-to-insulator transitions.\cite{Qazilbash2007,Zheng2015} In this section, we study a symmetric bond dissociation of a molecular vanadium oxide, \ce{V3O3}, which is spin-frustrated under \ce{D3_h} symmetry. It is probably not relevant to the strong correlations of \ce{VO2} in bulk, but we found this molecule interesting enough to study. Each \ce{V(II)} in a \ce{VO} unit has an electron configuration of d$^2$s$^1$ as opposed to the more commonly seen d$^3$ \insertnew{and the VO molecule has a X $^4\Sigma^-$ ground state}.\cite{Miliordos2007} 
This is not an artifact from approximate quantum chemistry models and was confirmed experimentally before.\cite{Kasai1968} We used \insertnew{a fixed} \ce{VO} bond length of $1.547431$ \AA{} throughout and obtained the PECs within the def2-SVP basis set \cite{Weigend2005} by varying the distance between \ce{V} and the center of the triangle.
The asymptote corresponds to three \ce{VO}(X$\:^4\Sigma^-$). 
\begin{figure}[h!]
\includegraphics[scale=0.5]{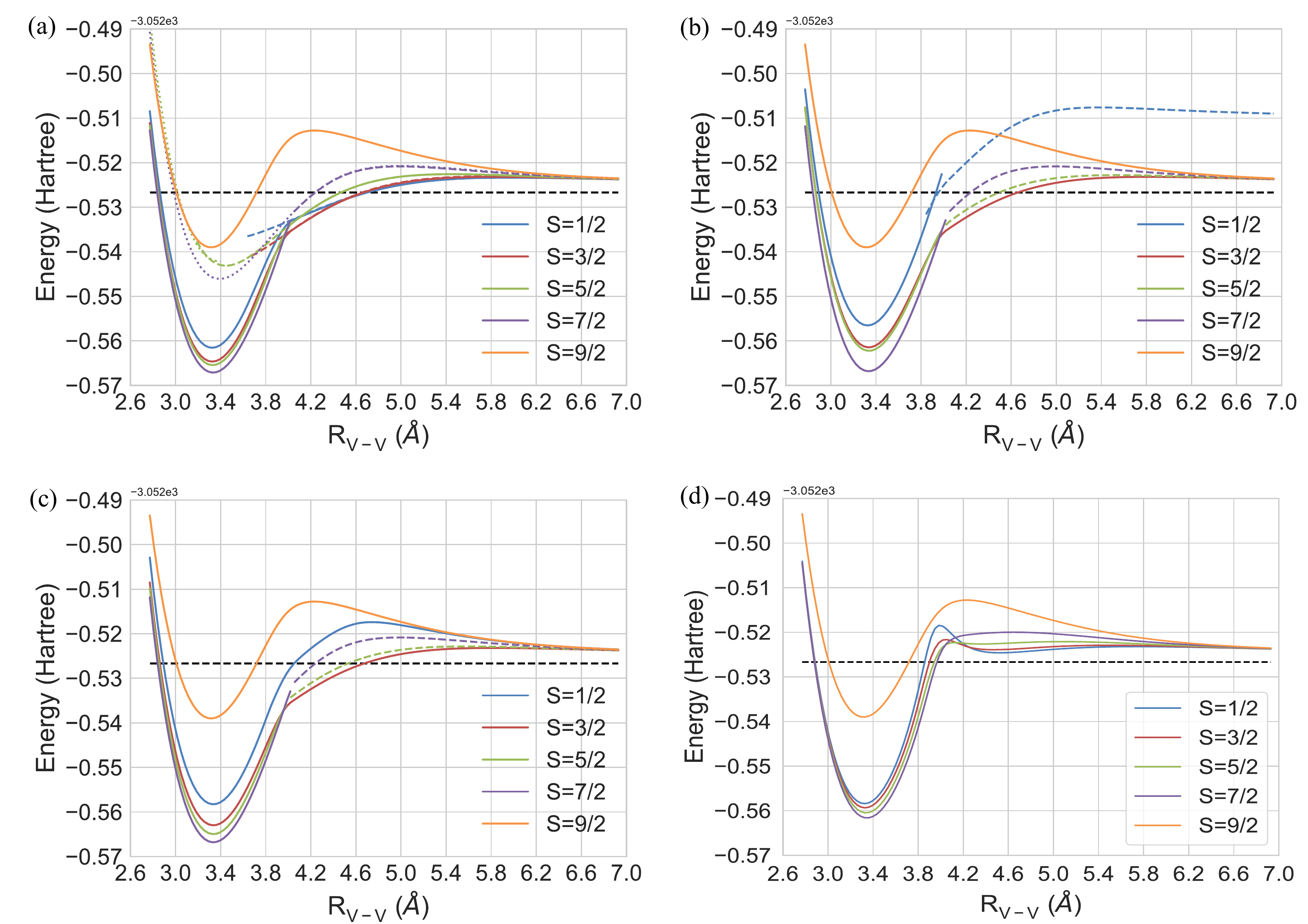}
\caption{\label{fig:v3o3pes}
Potential energy curves of a symmetric dissociation of \ce{V3O3} from
  {(a)} CASSCF (9e, 9o),
  {(b)} CCVB, 
  {(c)} CCVB+i3, \revinsert{and {(d)} SF-CAS.}
The black dotted line indicates the asymptotic energy of three \ce{VO}(X$\:^4\Sigma$), $E = -3052.5267114\: E_h$.
Different line styles for each spin state represent different local minima.
}
\end{figure}

Based on the CASSCF(9e, 9o) results, the ground state is ferromagnetic with $S=7/2$ within the active space. However, 
the CASSCF solutions other than that of $S=9/2$ all break spatial symmetry ($\text{D}_\text{3h}$) to some extent and this \insertnew{artificial symmetry breaking} indicates that the size of active space may not be fully appropriate \insertnew{with CASSCF orbital optimization}. This spatial symmetry breaking may also be the reason that the energy of $S=9/2$ is apparently too high compared to the rest. Moreover, the appearance of multiple solutions also indicates the same. This particular choice of the active space is made for the purpose of benchmarking like before, and it would be interesting to relax this pairing active space constraint in CASSCF and compare against CCVB-SD in the future. 

\insertnew{As for lower spin states of other ferromagnetic systems}, commonly used determinant based CI methods suffer from spin-contamination, so we employed the configuration state function (CSF) based CASSCF method in \texttt{Orca}. Some of the CASSCF results presented below started from CCVB or CCVB+i3 orbitals which are in general a very good guess. With determinant based CI methods, CASSCF can be very prone to high spin-contamination and often just collapses to an unwanted spin-state. \revreplace{This is quite common to observe}{This is commonly obeserved} when trying to obtain a low-spin state when the ground state is a high-spin state. We observed this quite frequently when starting from CCVB orbitals and therefore for those which used CCVB orbitals as a guess we added a penalty function to the electronic Hamiltonian to penalize the contaminants as implemented in \texttt{PySCF}, \insertnew{which is to add $\lambda (\langle \hat{S}^2 \rangle - S_z(S_z+1))^2 $ where $\lambda$ is a level-shift parameter}.

The CASSCF solutions in Figure \ref{fig:v3o3pes} (a) show quite interesting results. Near the minimum ($\sim 3.36$ \AA), the $S=1/2- 7/2$ states are very close in energy. Those states are all within a 7 kcal/mol energy window \insertnew{of each other}, and this is indeed the hallmark of SSC. Electrons are well localized and flipping one of the spins costs only a small energy penalty. We note that the system exhibits a strikingly slow convergence to the asymptote as the bond length increases. This slow algebraic decay is due to the fact that each X $^4\Sigma^-$ \ce{VO} is polar \insertnew{(roughly \ce{V+O-}\cite{Miliordos2007})}, and therefore the system exhibits multipolar interactions at long range. This has been verified by a log-log plot of energy-distance.

\insertnew{As is evident from} Figure \ref{fig:v3o3pes} (b), CCVB solutions capture all the qualitative features of their CASSCF counterparts except for the $S=1/2$ state at the dissociation limit. Similar to \ce{N3}, to describe the dissociation of the $S=1/2$ state one needs 3P substitutions. We observed multiple solutions in the case of the $S=3/2$ state similar to those obtained for \ce{N3}. Each of them corresponds to \insertnew{one of the} Lewis structures of $S=3/2$ described in Figure \ref{fig:n3lewis} (b) and (c), replacing N's with VO's. However, we only present the solution that dissociates properly. This solution is quite delocalized at $R=4.0$ \AA{} unlike the localized solution we found in \ce{N3}. The CCVB+i3 in Figure \ref{fig:v3o3pes} (c) shows only one solution in $S=1/2$ and it dissociates properly. The other states are more or less the same as those in CCVB. Both CCVB and CCVB+i3 correctly predict the relative energy ordering of different spin states near equilibrium.

\revinsert{
SF-CAS results are presented in Figure \ref{fig:v3o3pes} (d). These results are based on spatially symmetric wavefunctions. All the curves in Figure \ref{fig:v3o3pes} (d) are smooth unlike those obtained from CASSCF and CCVB methods. CASSCF, CCVB, and CCVB+i3 do not capture a small hump in $S=1/2$ present in SF-CAS. 
Instead, they exhibit a first-order derivative discontinuity due to the coexistence of two low-lying solutions. 
We believe that the discontinuity is closely related to the existence of the hump in the $S=1/2$ state of SF-CAS.
The SF-CAS relative energy ordering of different spin states near equilibrium agrees with other methods. It will be interesting to study these spin gaps in conjunction with dynamic correlation treatments to draw quantitative conclusions.
}

\begin{figure}[h!]
\includegraphics[scale=0.52]{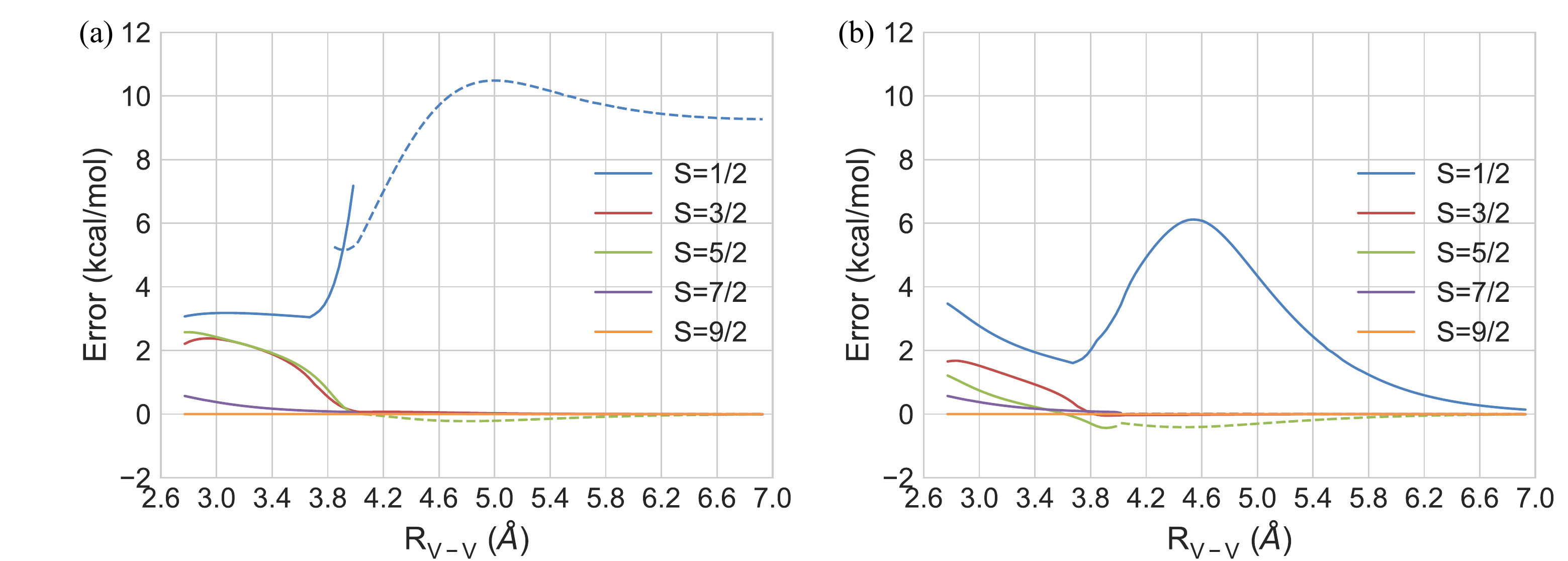}
\caption{\label{fig:v3o3error}
Errors in absolute energy relative to CASSCF of a symmetric dissociation of \ce{V3O3} for
  {(a)} CCVB and
  {(b)} CCVB+i3.
The line style of each line is consistent with that of Figure \ref{fig:v3o3pes}.
}
\end{figure}

The errors in absolute energy relative to CASSCF are shown in Figure \ref{fig:v3o3error} for both CCVB and CCVB+i3. All the errors are much smaller than those for \ce{N3}, which suggests that the bonding in \ce{V3O3} involves much smaller dynamic correlations compared to \ce{N3} within the active space. Clearly, CCVB+i3 shows improved results compared to CCVB, and the key improvement is the exact bond dissociation of $S=1/2$. It improves the $S=3/2$ and $S=5/2$ states by 1--2 kcal/mol. We also note that both CCVB and CCVB+i3 exhibit a slight non-variationality ($\le$0.50 kcal/mol) in $S=5/2$. The CASSCF calculations starting from those orbitals converged to higher values. As it is not significant in magnitude, we do not find it very concerning. 

The CCVB+i3 amplitudes of the $S=1/2$ state at $R = 4.0$ \AA{} are shown in Table \ref{tab:v3o3s2}. There are several qualitatively different features compared to the \ce{N3} results. 
The 2P amplitudes involving an OS pair localized on VO(1) and a CS pair connecting two other VO's are much larger. This is opposite to what was observed in \ce{N3}.
Usually, when 2P amplitudes are much larger than $1/\sqrt{3}$ in magnitude, it is often possible to identify a different PP reference. In this case, it is not obvious to us if there exists a better reference. The 3P amplitudes show two $t_{KL\mu}$ amplitudes that are larger than the largest $t_{KLM}$ amplitudes. In the case of \ce{N3}, the largest $t_{KL\mu}$ amplitudes were smaller than the largest $t_{KLM}$ amplitudes. 
However, we confirm that the condition we found for having significant $t_{KLa}$ still holds.
\begin{table}
  \centering
\begin{tabular}{rr|r}\hline
$K$&\multicolumn{1}{c}{$a$}\vline&\multicolumn{1}{c}{$t_{Ka}$}\\
\hline
2    &     5    &   -1.32106\\
3    &     5    &   -1.31987\\
4    &     5    &    0.57741\\
2    &     3    &    0.57734\\
3    &     4    &    0.57708\\
2    &     4    &    0.57661\\
1    &     5    &   -0.57650\\
1    &     3    &    0.57425\\
1    &     2    &    0.57323\\
1    &     4    &   -0.57174\\
\hline
\end{tabular}
  \quad
  \begin{tabular}{rrr|r}\hline
$K$ & \multicolumn{1}{c}{$L$} & \multicolumn{1}{c}{$a$}\vline & \multicolumn{1}{c}{$t_{KLa}$} \\
\hline
     2     &    4     &    5     &   1.34229\\
     3     &    4     &    5     &   1.34158\\
     1     &    3     &    4     &  -0.80939\\
     1     &    2     &    4     &  -0.80723\\
     1     &    2     &    5     &   0.52392\\
     1     &    3     &    5     &   0.52209\\
     1     &    4     &    5     &   0.00045\\
     1     &    2     &    3     &  -0.00045\\
     2     &    3     &    5     &  -0.00026\\
     2     &    3     &    4     &  -0.00015\\
\hline
  \end{tabular}
  \caption{The CCVB+i3 2-pair and 3-pair amplitudes of \ce{V3O3} ($S=1/2$) at $R=4.0$ \AA.
  1--4 are CS pairs and 5 is an OS pair. The CS pair 1 consists of two s-like orbitals, the CS pair 2 consists of a d-like orbital and a s-like orbital, and the rest contains only d-like orbitals. Pair labels are consistent with those in Figure \ref{fig:n3lewis} (a) where each \ce{N} is replaced by a \ce{VO}.
  }
  \label{tab:v3o3s2}
\end{table}

When the $S=5/2$ state at $R=4.0$ \AA{} is evaluated with localized orbitals (from the $S=1/2$ state and unpairing polarized pairs), the 3P amplitudes are significant. The same analysis for \ce{N3} revealed that the 3P amplitudes are negligible, so this result in \ce{V3O3} is quite different although the geometry setup is the same.
We suspect that this is because in \ce{V3O3} there may be more than one way to \revreplace{spin-coupled}{spin-couple} high-spin fragments to correctly reach the asymptote and some of them do not need the 3P substitutions and some do. 
If the former is the case, it is not too surprising that the energy difference between CCVB and CCVB+i3 is about 0.01 kcal/mol.
It is, however, possible that some higher-body correlations functions beyond two-body correlators (or 4-point correlators) will show larger differences between different spin-couplings (i.e., those that involve 3P substitutions and those that do not). Those higher-body correlation functions such as three-body Green's functions are often studied in nuclear physics.\cite{Ethofer1969,Bender1988}
After orbital optimization, \insertnew{the $S=5/2$} CCVB orbitals become delocalized and the 3P amplitudes become negligible.

\subsection{The [\ce{CaMn3O4}] Subunit of Oxygen-Evolving Complex}
\begin{figure}[h!]
\includegraphics[scale=0.5]{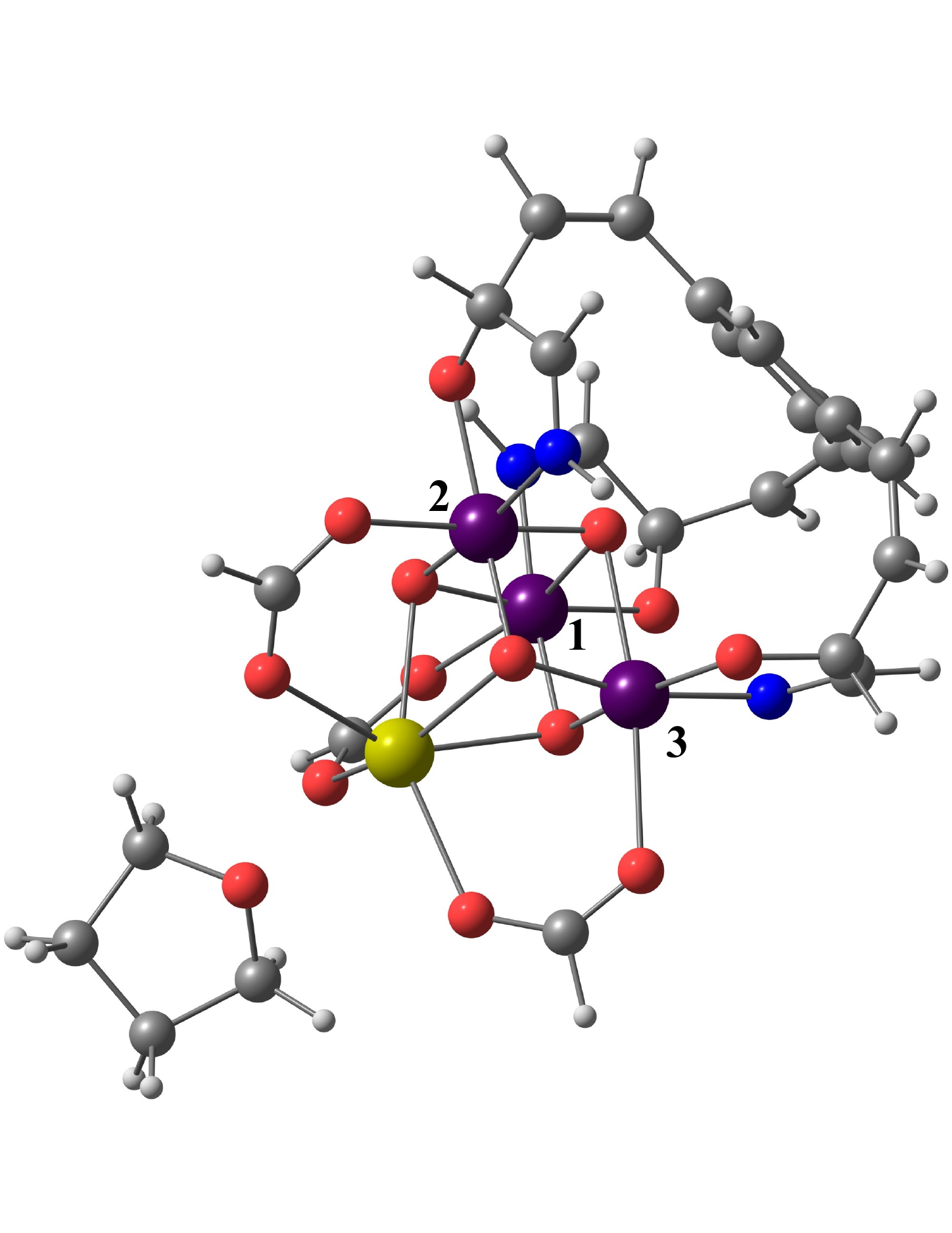}
\caption{\label{fig:cubane}
The molecular structure of a simplified synthetic model of the cubane subunit of OEC. The color code we used is as follows: grey: \ce{C}, white: \ce{H}, red: \ce{O}, blue: \ce{N}, purple: \ce{Mn}, and yellow: \ce{Ca}.
The numbers indicate the labels for Mn atoms.
The distances between Mn atoms are as follow:
\ce{Mn(1)}-\ce{Mn(2)} = 2.83 \AA,
\ce{Mn(2)}-\ce{Mn(3)} = 2.90 \AA, and
\ce{Mn(3)}-\ce{Mn(1)} = 2.95 \AA.
}
\end{figure}
The cubane subunit, [\ce{CaMn3O4}], in the oxygen-evolving complex (OEC) has attracted a lot of interest in both experimental and theoretical chemistry.\cite{Mukhopadhyay2004,Lundberg2004,McEvoy2006,Fliegl2009,Dismukes2009,Kanady2011,Ames2011,Mukherjee2012,Kanady2013,Kurashige2013,Lee2015} From a theoretical chemistry point of view, this is a very challenging system because it requires a balanced treatment of both static and dynamic correlations and there is no readily usable quantum chemistry model which can achieve this. 
Moreover, understanding electron correlations in this molecule \insertnew{may assist in the} rational design of synthetic OECs, \insertnew{which is also of much interest}.\cite{Kanady2011,Mukherjee2012,Zhang2015}

Most recent studies have focused on studying the complete complex including the fourth ``dangling'' \ce{Mn} atom. 
The presence of the fourth \ce{Mn} reduces the effect of 3P amplitudes significantly in the ground state, and hence CCVB \insertnew{appears to be} well-suited for the full complex. For \insertnew{a more demanding test of} CCVB+i3, we therefore considered the cubane subunit without the dangling \ce{Mn}, which shows significant 3P contributions as we shall see. 

The geometry shown in Figure \ref{fig:cubane} was taken from ref. \citenum{Lee2015}, which is a simplified cubane subunit of the synthetic model reported by Agapie and co-workers.\cite{Kanady2011} The structure is very similar to what Agapie and co-workers used in their theoretical study \cite{Kanady2013} and the corresponding molecule in ref. \citenum{Kanady2013} was neutral with [\ce{Ca}Mn$^\text{IV}_3$\ce{O4}]. However, the oxidation state studied in ref. \citenum{Lee2015} is [\ce{CaMn}$^\text{III}_2$Mn$^\text{IV}$\ce{O4}] with overall $-2$ charge. Since we are interested in maximizing spin-frustration, we chose to study [\ce{CaMn}$^\text{IV}_3$\ce{O4}] with charge neutrality. The natural active space is (9e, 9o) which includes all the d-electrons in 3 \ce{Mn}$^\text{IV}$ (d$^3$). This active space is small enough that exact CASSCF can easily be performed. This active space may be too small to describe the system properly, but it includes every d-orbital with strong open-shell character which are the primary source of strong correlation.

An interesting feature of this molecule is that the three \ce{Mn} atoms form a nearly perfect triangle and thus 3P configurations ought to play a crucial role as we learned from the isoelectronic model systems above. We used the def2-SVP basis set \cite{Weigend2005} for hydrogen and carbon atoms and the def2-TZVP basis set \cite{Weigend2005} for everything else. We also employed the density-fitting approximation to the two-electron integrals with the corresponding density-fitting bases.\cite{Hattig2005}

We employed several computational approaches to compute spin gaps and compared against exact CASSCF as shown in Table \ref{tab:cubane}. The CASSCF results show a monotonic increase in the spin gap as we go from the high-spin state to the low-spin state, indicating that the complex is ferromagnetic within the active space employed here. 
Remarkably, all the spin states lie within 1 kcal/mol and this is again a signature of SSC. 

SF-CAS energies are also computed with the $S=9/2$ reference, and it should be the upper bound to the CASSCF energies for each spin state as both methods are variational. The SF-CAS energies show the same trend as CASSCF and it is qualitatively accurate for this system. As orbitals are very well localized, the effect of orbital optimization is expected to be small, and this is consistent with the observation here. The perturbative correction to the SF-CAS states (SF-CAS(h,p)$_1$) \cite{Mayhall2014}, which attempts to incorporate orbital relaxation, fails quite significantly, and it yields negative spin gaps. \insertnew{This suggests that orbital relaxation is not negligible and one may try to regularize the energy denominators.\cite{Lee2018}}

CCVB and CCVB+i3 predict the relative spin gaps of $S=3/2$ and $S=5/2$ wrong, but essentially the errors are all within 1 kcal/mol for those states. 
\revinsert{These incorrect orderings may well be fixed by a full CC model which generalizes these CCVB models. 
Such a full CC model incorporates ionic excitations and thus will provide more accurate energies within this active space.}
CCVB fails catastrophically to describe the $S=1/2$ state yielding a spin gap of roughly 29 kcal/mol! In contrast, CCVB+i3 yields a quantitatively accurate result. This highlights the role of 3P substitutions in spin-frustrated systems.
\begin{table}
  \centering
\begin{tabular}{r|r|r|r|r|r}\hline
\multicolumn{1}{c}{S}\vline
& \multicolumn{1}{c}{CASSCF}\vline 
& \multicolumn{1}{c}{SF-CAS} \vline
& \multicolumn{1}{c}{SF-CAS(h,p)$_1$} \vline
& \multicolumn{1}{c}{CCVB}\vline
&\multicolumn{1}{c}{CCVB+i3}\\
 \hline
1/2 & 0.853 & 1.074 & N/A & 29.816 & 1.412\\
3/2 & 0.662 & 0.884 & -0.012 & 0.791 & 0.796\\
5/2 & 0.535 & 0.672 & -0.011 & 0.889 & 0.889\\
7/2 & 0.327 & 0.377 & -0.006 & 0.588 & 0.588\\
9/2 & 0.000 & 0.000 & 0.000 & 0.000 & 0.000\\
\hline
\end{tabular}
  \caption{Relative energies (kcal/mol) of different spin states from different methods.
  N/A means ``not available'' due to the limited computational resource.
  }
  \label{tab:cubane}
\end{table}

Turning to the 2P and 3P amplitudes of CCVB+i3 of $S=1/2$ in Table \ref{tab:cubane2}, we observe qualitatively similar results to those for \ce{N3}. Most of the 2P amplitudes are close to $1/\sqrt{3}$ in magnitude and only those types of $t_{K\mu}$ that were small in Table \ref{tab:n3s2} are negligible here. 

The natural orbital occupation numbers (NOONs) from CASSCF show strong open-shell characters in all 9 orbitals regardless of the spin state (i.e., they are all near 1.0). Both CCVB and CCVB+i3 successfully capture this (i.e., NOONs are all near 1.0). It is interesting that the NOONs of the $S=1/2$ state in CCVB are almost the same as those of CASSCF even though its energy is 29 kcal/mol higher. In this particular case, orbital relaxation upon going from CCVB orbitals to CCVB+i3 orbitals is negligible, \insertnew{which indicates that} including the proper 3P spin-coupling vectors is crucial to obtain an accurate energy. Reading CCVB orbitals into CCVB+i3 yields an energy that is higher only by 1 mH than the optimized energy.
\begin{table}
  \centering
\begin{tabular}{rr|r}\hline
$K$&\multicolumn{1}{c}{$a$}\vline&\multicolumn{1}{c}{$t_{Ka}$}\\
\hline
     1     &    5     &   0.57570\\
     4     &    5     &   0.57564\\
     3     &    2     &   0.57497\\
     1     &    4     &   0.57201\\
     3     &    1     &   0.57116\\
     2     &    1     &   0.57102\\
     2     &    4     &  -0.57029\\
     3     &    4     &  -0.56979\\
     2     &    5     &  -0.13906\\
     3     &    5     &  -0.13819\\
\hline
\end{tabular}
  \quad
  \begin{tabular}{rrr|r}\hline
$K$ & \multicolumn{1}{c}{$L$} & \multicolumn{1}{c}{$a$}\vline & \multicolumn{1}{c}{$t_{KLa}$} \\
\hline
1     &    3     &    4    &    0.80176\\
1     &    2     &    4    &    0.80114\\
1     &    2     &    5    &    0.50033\\
1     &    3     &    5    &    0.49939\\
3     &    4     &    5    &   -0.30389\\
2     &    4     &    5    &   -0.30333\\
2     &    3     &    4    &   -0.00029\\
2     &    3     &    5    &   -0.00023\\
1     &    2     &    3    &   -0.00019\\
1     &    4     &    5    &   -0.00011\\
\hline
  \end{tabular}
  \caption{The CCVB+i3 2-pair and 3-pair amplitudes of the cubane subunit for $S=1/2$.
  1--4 are CS pairs and 5 is an OS pair. Pair labels are consistent with those in Figure \ref{fig:n3lewis} (a) where each \ce{N} is replaced by a \ce{Mn} atom.
  }
  \label{tab:cubane2}
\end{table}

\subsection{\ce{P5} ($\text{D}_\text{5h}$) -- Spin-frustration in a pentagon}
Clusters of phosphorus have been studied theoretically and experimentally by many researchers.\cite{Jones1990,Haeser1992,Huang1995,Bcker1995,HUANG1996,Chen1999,Chen2000,Chen2000a,Bulgakov2004} 
Here, we studied \ce{P5} and fixed its geometry to $\text{D}_\text{5h}$ so that the molecule is forced to be spin-frustrated. It is an interesting spin-frustrated model system that is beyond the triangular geometric frustrations that have been discussed in this work so far. UHF can dissociate properly \insertnew{to quartet P atoms} only when $M_S=3/2, 9/2, 15/2$.

The natural choice of an active space is (15e, 15o) \insertnew{for which} exact CASSCF is demanding so we used the recently developed selected CI method, heat-bath CI (HCI) with orbital optimization (HCISCF) \cite{Smith2017} for the reference benchmark data.
We compare variational HCISCF energies and CCVB energies. 
In HCISCF, we test different values of $\epsilon_1$ which controls the number of determinants included in the variational space.
The smaller $\epsilon_1$ value yields the larger variational space and thus the result becomes more accurate.
As our focus is the strong correlation in this system, we chose a relatively stretched geometry, $R_\text{P-P}$ = 4.1145 \AA.
\revinsert{This stretched geometry makes CCVB methodologies particularly well-suited since ionic excitations are negligible.
The only important excitations are of the spin-flip type.}
We employed the def2-SVP basis set throughout.\cite{Weigend2005}
\begin{table}
  \centering
  \begin{tabular}{r|c|r|r|r|r|r}\hline
\multicolumn{1}{c}{$S$}\vline 
& \multicolumn{1}{c}{CCVB} \vline
& \multicolumn{1}{c}{CCVB+i3} \vline
& \shortstack{HCISCF\\($\epsilon_1=10^{-3}$)}
& \shortstack{HCISCF\\($\epsilon_1=10^{-4}$)}
& \shortstack{HCISCF\\($\epsilon_1=10^{-5}$)}
& \shortstack{HCISCF\\($\epsilon_1=10^{-6}$)}\\
\hline
1/2& 7.14 (7.14)&-3.22&173.50&-4.78&-5.35&-5.37\\
3/2& 5.92 (5.92)&-2.98&85.54&-4.52&-5.04&-5.05\\
5/2& 4.59 (4.59)&-2.68&131.31&-4.01&-4.60&-4.60\\
7/2& 10.92 (-3.85)&-3.85&-0.89&-3.39&-3.95&-3.96\\
9/2& -2.18 (-2.18)&-2.20&-1.24&-2.70&-3.18&-3.18\\
11/2& -1.00 (-1.79)&-1.79&-1.35&-2.20&-2.26&-2.26\\
13/2& -0.48 (-0.48)&-0.48&-0.99&-1.19&-1.20&-1.20\\
15/2& 0.00 (0.00)&0.00&0.00&0.00&0.00&0.00\\
\hline
  \end{tabular}
  \caption{
The relative energies (kcal/mol) of different spin states of \ce{P5}.
The CCVB energies in parentheses are from the CCVB solutions where converged CCVB+i3 orbitals were used as an initial guess.
\revinsert{For $S=15/2$, every method presented here is exact since ROHF is exact for that state.
The corresponding $S=15/2$ ROHF energy is -1702.98131 $E_h$.
These spin-gaps are directly comparable across different methods as they are measured with respect to this same energy.
}
  }
  \label{tab:p5energy}
\end{table}

There can be more than one solution for each spin state as we saw in the previous examples. Here, we focus on CCVB solutions from the ``spin-ladder guess'' procedure described before. 
In other words, all the CCVB and CCVB+i3 calculations are performed using the orbitals from $S=1/2$ as a guess. For HCISCF calculations, the CCVB+i3 converged orbitals were used as an initial guess. 

In Table \ref{tab:p5energy}, we present the spin gaps of CCVB and CCVB+i3 along with the HCISCF spin gaps as a reference.
Comparing the CCVB and CCVB+i3 energies, significant energy differences are observed for $S=1/2,3/2,5/2,7/2$. 
As UHF can dissociate $M_S=3/2$ properly, there must be a CCVB solution for $S=3/2$ that is lower in energy than what we found here. 
However, we discuss only those obtained from an $S=1/2$ initial guess for the purpose of demonstration. 
For $S=7/2$, CCVB yields a localized solution with a quite high energy while CCVB+i3 yields a delocalized solution with a lower energy and no significant 3P amplitudes. 
By reading the CCVB+i3 orbitals into CCVB, we were able to obtain a CCVB solution that reaches the correct asymptote.
HCISCF energies are converged at around the $\epsilon_1$ value of $10^{-5}$ and HCISCF of $\epsilon_1=10^{-3}$ shows \insertnew{very unconverged} energies. In terms of energies, CCVB+i3 lies between HCISCF of $\epsilon_1=10^{-3}$ and $\epsilon_1=10^{-4}$, \insertnew{although the nature of the errors is quite different.}

\begin{table}
  \centering
  \begin{tabular}{r|r|r|r|r|r}\hline
\multicolumn{1}{c}{$S$}\vline 
& \multicolumn{1}{c}{CCVB+i3} \vline
& \shortstack{HCISCF\\($\epsilon_1=10^{-3}$)}
& \shortstack{HCISCF\\($\epsilon_1=10^{-4}$)}
& \shortstack{HCISCF\\($\epsilon_1=10^{-5}$)}
& \shortstack{HCISCF\\($\epsilon_1=10^{-6}$)}\\
\hline
1/2 & 35 & 942 & 5331 & 145087 & 488263\\
3/2 & 39 & 885 & 6324 & 143864 & 465000\\
5/2 & 40 & 552 & 4946 & 96852 & 237935\\
7/2 & 38 & 2650 & 32849 & 158377 & 330978\\
9/2 & 33 & 772 & 4383 & 19184 & 35882\\
11/2 & 25 & 68 & 1181 & 2743 & 3538\\
13/2 & 14 & 31 & 87 & 122 & 122\\
15/2 & 0 & 0 & 0 & 0 & 0\\
\hline
  \end{tabular}
  \caption{
The number of independent wavefunction parameters used in each method in \ce{P5}. For CCVB+i3, this number is the same as \insertnew{for CCVB: 2-pair} amplitudes $t_{Ka}$ plus the number of polarization angles $\theta_K$. For HCISCF, this is the number of determinants minus one due to the wavefunction normalization.
  }
  \label{tab:p5params}
\end{table}

In Table \ref{tab:p5params}, we present the number of independent wavefunction parameters used in each method. \insertnew{An advantage} of CC methods is the ability to describe chemical systems with a much more compact representation through a cluster expansion, which linear CI wavefunctions do not offer. Table \ref{tab:p5params} shows that the number of parameters in the CCVB wavefunction is much smaller than in HCISCF. Comparing CCVB and HCISCF of $\epsilon_1=10^{-3}$ or $\epsilon_1=10^{-4}$ which are similar in accuracy, we see that CCVB has 20-150 times fewer parameters than HCISCF for $S=1/2$. Remarkably, for $S=7/2$ CCVB is more accurate than HCISCF of $\epsilon_1=10^{-4}$ while involving roughly 860 times fewer parameters.
\begin{figure}[h!]
\includegraphics[scale=0.4]{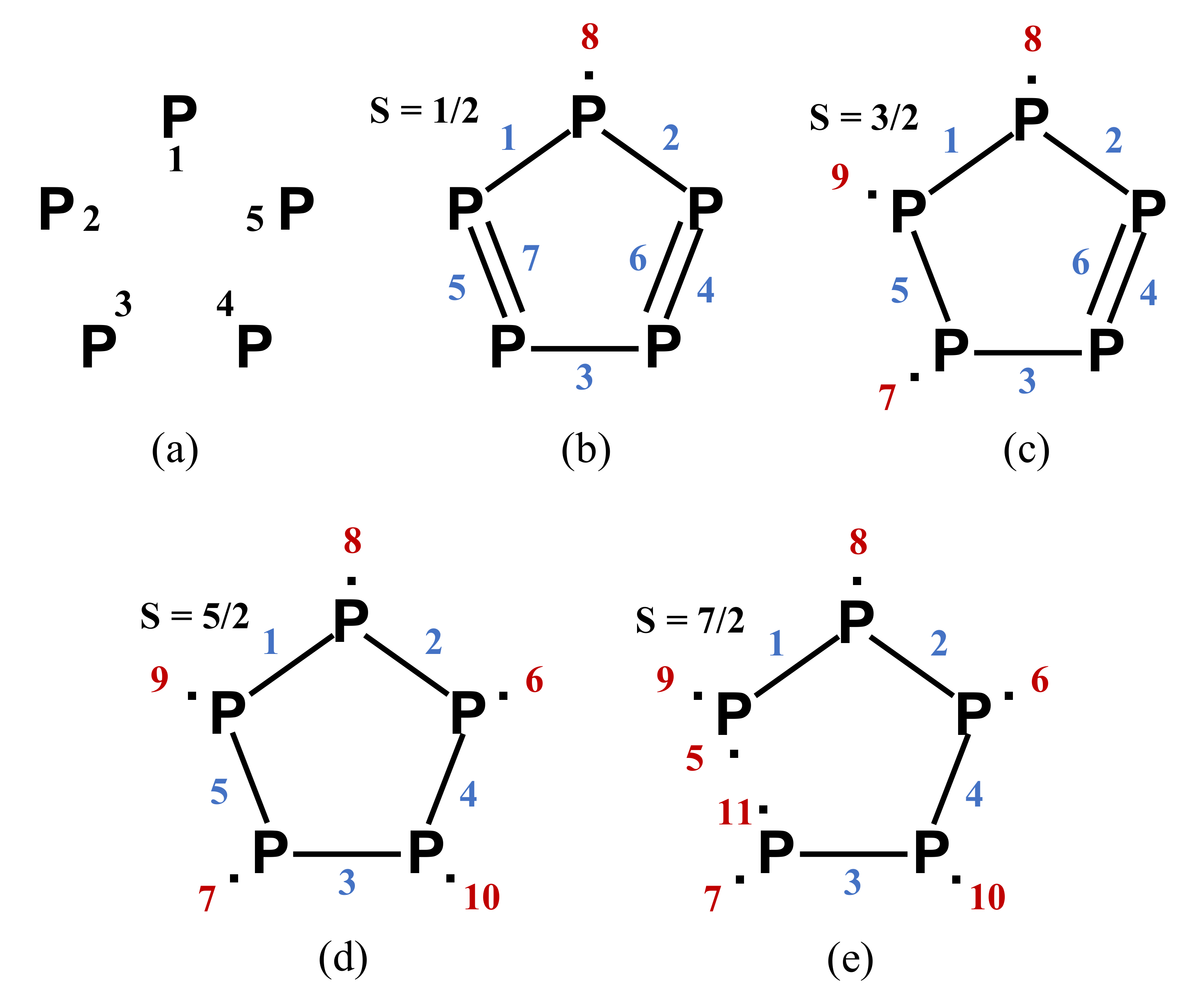}
\caption{\label{fig:p5lewis}
Panel (a) shows how we label five phosphorus atoms.
The rest of panels illustrate represent possible Lewis structures of \ce{P5} for each spin state:
(b) $S=1/2$, (c) $S=3/2$, (d) $S=5/2$, and (e) $S=7/2$.
The number next to each bond or unpaired electron is used to label a CS (blue) or OS (red) pair.
Both of CCVB methods yield localized solutions for (b) to (d), and only CCVB yields a localized solution for (e).
Every P-P bond consists of p-orbitals and each P atom has three p-orbitals which yields an active space of (15e, 15o).
}
\end{figure}

Lastly, we discuss \insertnew{3-pair (3P)} amplitudes along with corresponding Lewis structures shown in Figure \ref{fig:p5lewis}. In the previous examples ($\text{D}_\text{3h}$), we made empirical observations on significant 3P amplitudes. We will see how those \insertnew{transfer to this 15-electron} $\text{D}_\text{5h}$ case. Because there are quite a few 3P amplitudes, we will visualize a subset of those amplitudes as opposed to presenting every one of them. One way to visualize 3P amplitudes is to fix one of the three indices and look at the matrix indexed by the other two indices. We will discuss such matrices below.

\begin{figure}[h!]
\includegraphics[scale=0.5]{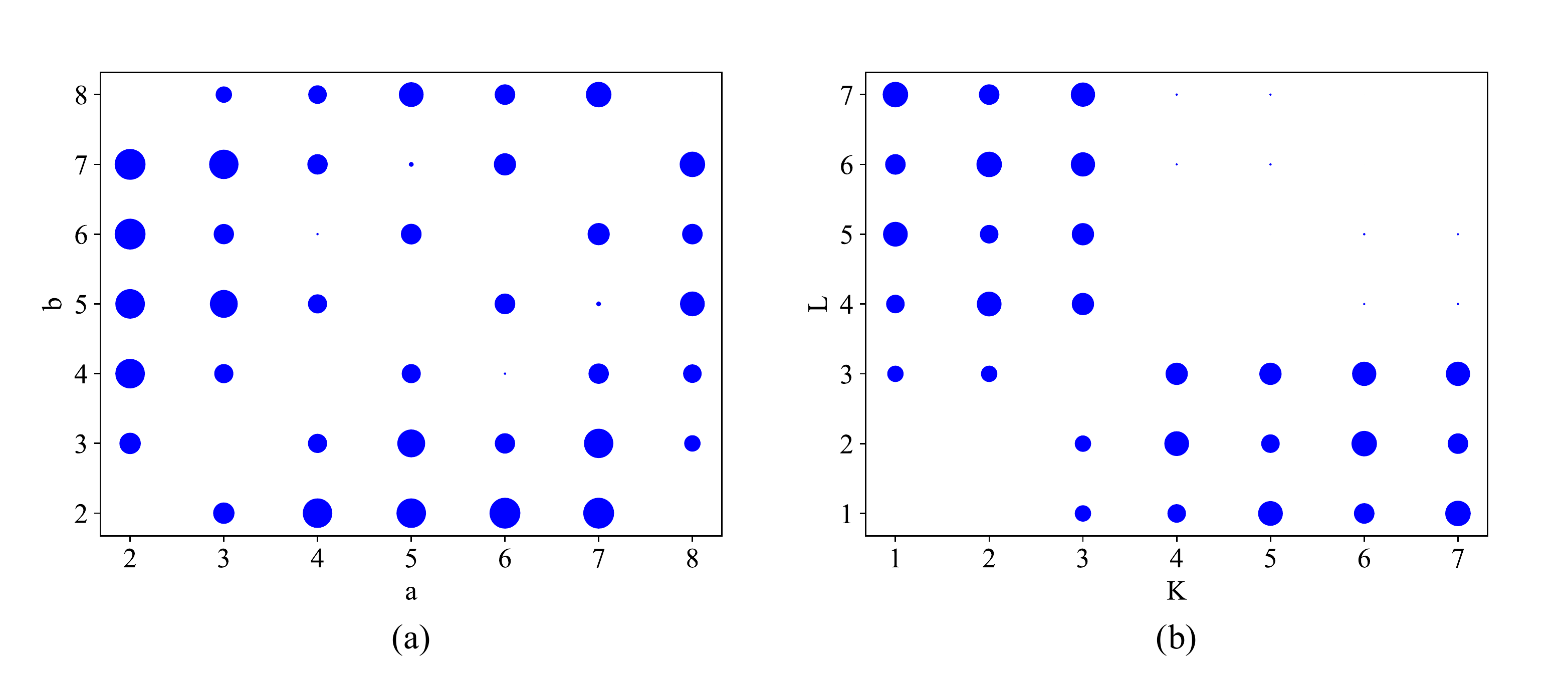}
\caption{\label{fig:p53p2}
The 3P amplitudes of the $S=1/2$ state of \ce{P5}. (a) $t_{Kab}$ when $K = 1$, (b) $t_{KL\lambda}$ when $\lambda = 8$. The labeling here is consistent with the localized orbitals and labels in Figure \ref{fig:p5lewis} (b). Pairs 1-7 are CS and pair 8 is OS.
\revreplace{The largest circle}{The circle of largest area} represents an amplitude of magnitude 0.18533 in (a) and an amplitude of magnitude 0.27933 in (b).
}
\end{figure}

Figure \ref{fig:p53p2} shows the 3P amplitudes of $S=1/2$. Panel (a) and (b) fix a CS pair and an OS pair, respectively. As is clearly shown, involving an OS index yields sparser 3P amplitudes. There are many significant $t_{KLM}$'s that involve pair 1 and 2. Basically, $t_{12M}$ for any CS pair $M$ is non-negligible. As seen in Figure \ref{fig:p5lewis} (b), those pairs form incomplete triangles. 
All of them involve more than three P atoms. Obviously, this could not be observed in the $\text{D}_\text{3h}$ examples since there are only three sites.
Panel (b) shows negligible $t_{KL8}$ when both $K$ and $L$ belong to $\{4,5,6,7\}$. Non-negligible amplitudes involve either pair 1, 2 or 3. Involving pair 1 or 2 is consistent with what we found in the \text{D}$_\text{3h}$ examples. Having more spin-frustrated sites gives a more rich spectrum of non-negligible 3P amplitudes and this is a manifestation of complex overall spin-coupling vectors whose total underlying dimension scales formally exponentially with the number of electrons.

\begin{figure}[h!]
\includegraphics[scale=0.5]{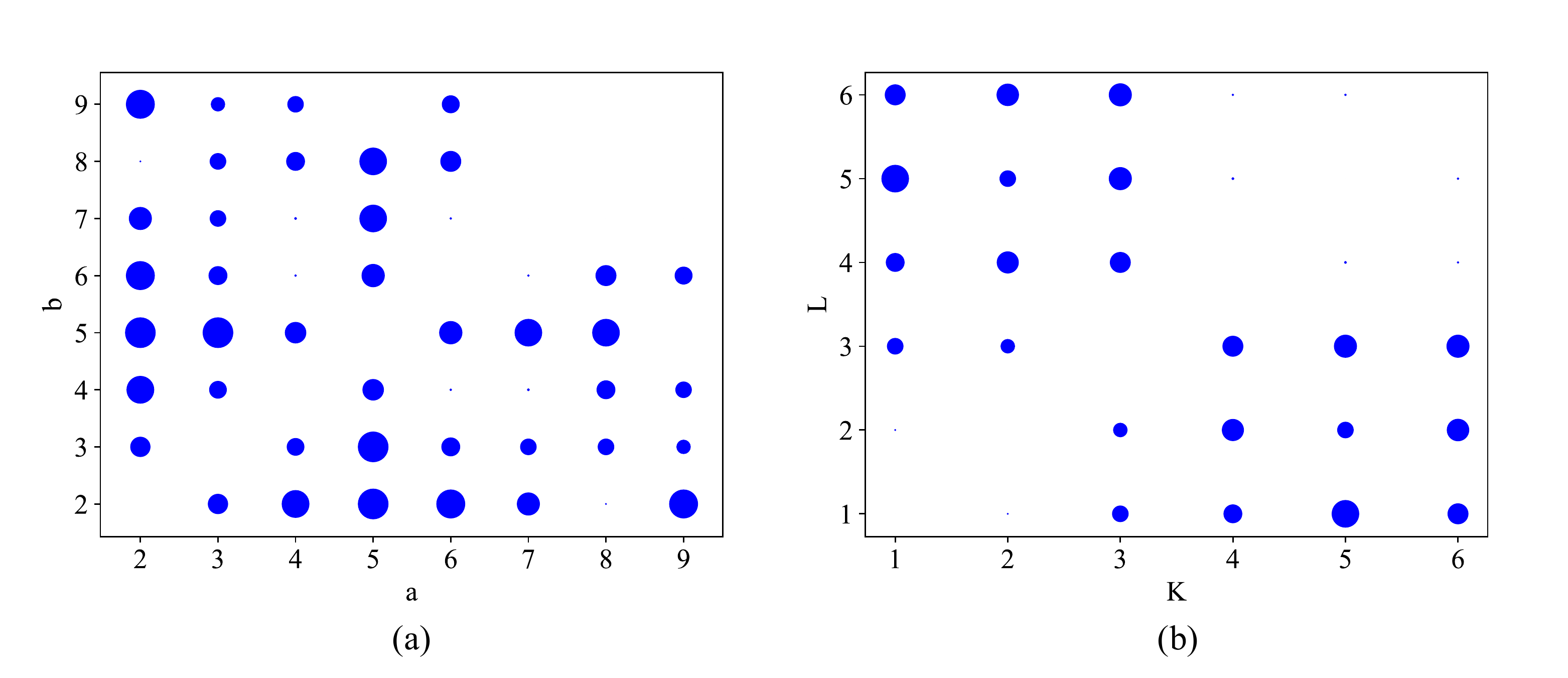}
\caption{\label{fig:p53p4}
The 3P amplitudes of the $S=3/2$ state of \ce{P5}. (a) $t_{Kab}$ when $K = 1$, (b) $t_{KL\lambda}$ when $\lambda = 8$. The labeling here is consistent with the localized orbitals and labels in Figure \ref{fig:p5lewis} (c). Pairs 1-6 are CS and pairs 7-9 are OS.
\revreplace{The largest circle}{The circle of largest area} represents an amplitude of magnitude 0.27484 in (a) and an amplitude of magnitude 0.21984 in (b).
}
\end{figure}

Figure \ref{fig:p53p4} illustrates the 3P amplitudes of the $S=3/2$ state. Panel (b) may be easy to understand because it is basically the same as a subblock of panel (b) in Figure \ref{fig:p53p2}. Unpairing a pair in $S=1/2$ basically results in removing a CS pair column and row in $t_{KL\lambda}$ for a given $\lambda$. At this stretched geometry, orbitals barely change, and hence it is not surprising to see this similarity in $t_{KL\lambda}$ across different spin states. Panel (a) exhibits a similar result in that its subblock (2 to 6) is very similar to that of $S=1/2$. The additional OS pairs 7 and 9 have non-negligible $t_{1L\mu}$. The other spin states, $S=5/2,7/2$, which correspond to (d) and (e) in Figure \ref{fig:p5lewis}, show more or less the same result. 

In summary, we have seen that \ce{P5} exhibits a much more complex spin-coupling pattern compared to those of the $\text{D}_\text{3h}$ examples. Namely, \ce{P5} shows many non-negligible $t_{KLa}$ amplitudes that are relevant in reaching the correct asymptote. Compared to HCISCF, it involves far fewer parameters, and yet their energies are comparable.

\subsection{The [\ce{Cr9}] Single Molecular Magnet}
Single molecular magnets (SMMs) have received a lot of interest lately since they can play a role of magnetic memory and potentially be used to build a quantum computer.\cite{Tejada2001,Leuenberger2001} A lot of theoretical studies on multinuclear complexes have been focused on broken-symmetry density functional theory (BS-DFT) often combined with the Heisenberg model to obtain a spin-spin coupling, $J$, between neighboring sites.\cite{Yamaguchi1986,Noodleman1986,Noodleman1992,Yamanaka1994,Baker2012} 

Along this line, Mayhall and Head-Gordon devised a simple and useful scheme that utilizes a single spin-flip wavefunction, maps the wavefunction to the Heisenberg model, and \revreplace{compute}{computes} spin-spin couplings between sites.\cite{Mayhall2014b,Mayhall2015} As long as the Heisenberg model is valid, their scheme is also valid. However, in practice it is hard to know whether the Heisenberg model is valid for a given system. Besides this method does not yield ab-initio wavefunctions for each spin state. This is a good motivation to try other alternatives. When the Heisenberg model is valid, CCVB methodologies become very powerful because there is no need for ionic excitations. It also targets each spin state in a state-specific way with orbital optimizations and yields spin-pure wavefunctions.
\begin{figure}[h!]
\includegraphics[scale=0.45]{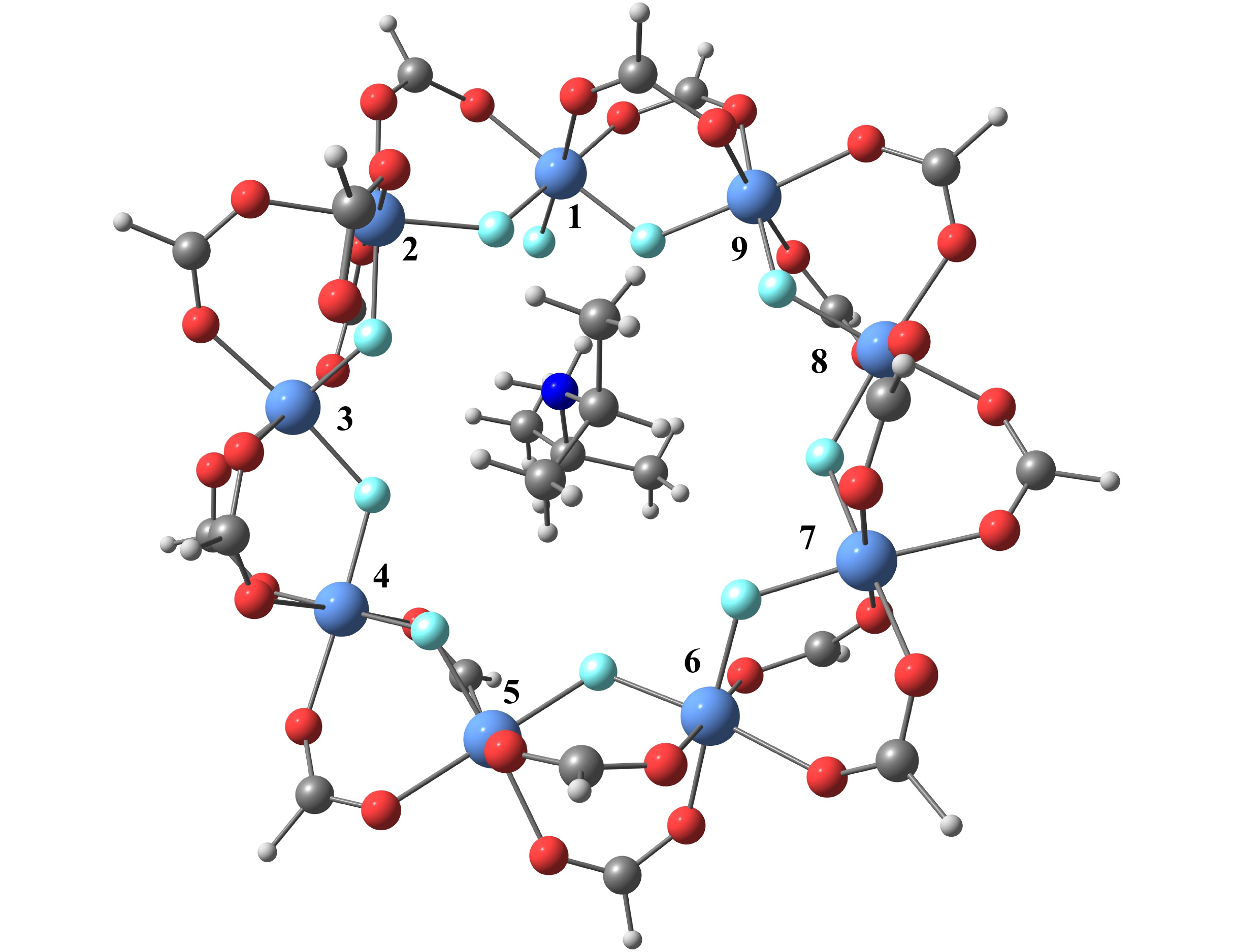}
\caption{\label{fig:cr9}
The molecular structure of the [\ce{Cr9}] SMM.
The color code we used is as follows: grey: \ce{C}, white: \ce{H}, red: \ce{O}, blue: \ce{N}, light blue: \ce{F}, and grey blue: \ce{Cr}.
The numbers next to Cr's indicate the label for each Cr. The molecule at the center is necessary to keep the system overall neutral and also found in the experimental crystal structure.
}
\end{figure}

In what follows, we discuss the electron correlation in the [\ce{Cr9}] SMM. The geometry shown in Figure \ref{fig:cr9} is taken from the crystal structure reported in ref. \citenum{Baker2012} (denoted as structure 4 therein), and all acetates are replaced with formates for the sake of computational simplicity. 
The \ce{C-H} bond length in the formates was adjusted to be 1.09 \AA. The neighboring \ce{Cr-Cr} distances range from 3.50 \AA{} to 3.90 \AA. The diisopropyl-ammonium ion in the center keeps the system overall neutral. 
\insertnew{This [\ce{Cr9}] SMM exhibits spin-frustation at equilibrium.}
All the chromiums are \ce{Cr}({III}) and have a $\text{d}$$^3$ electron configuration. Therefore, the natural choice of an active space in this system is (27e, 27o). This is currently beyond the scope of exact CASSCF.

Hence, we benchmarked CCVB against HCI. HCI is not orbital-invariant, so choosing the right set of orbitals is quite crucial. We took converged CCVB orbitals and ran HCI with those orbitals. 
Ideally, optimizing orbitals with HCI should produce the best benchmark numbers. Using CCVB orbitals as an initial orbital, we attempted orbital optimization with various values of $\epsilon_1$ in HCI. The orbital optimization exhibited an energy fluctuation of 10 $\mu$H, and we were not able to converge tightly. Therefore, we do not report those numbers and focus on the HCI results performed with CCVB orbitals (denoted as HCI//CCVB) for the following discussion. We used the def2-TZVP basis set \cite{Weigend2005} on Cr and the def2-SVP basis set \cite{Weigend2005} on all the other atoms along with the corresponding density-fitting bases.\cite{Hattig2005} In passing, we note that the value of using CCVB or PP orbitals for a subsequent HCI calculation has recently been pointed out by Zimmerman in the context of iFCI.\cite{Zimmerman2017a}

The amplitude equation of CCVB+i3 becomes somewhat ill-defined in this case. In other words, the Jacobian in Eq. \eqref{eq:squaresum} becomes nearly singular and thus finding solutions becomes extremely challenging. As mentioned earlier in Computational Details, we loosened the convergence threshold of the $t$-amplitudes to $10^{-8}-10^{-10}$ and did not perform orbital optimization for CCVB+i3. Instead, we performed CCVB+i3 calculations on converged CCVB orbitals. We denote this as CCVB+i3//CCVB in the following discussion. 

\begin{figure}[h!]
\includegraphics[scale=0.35]{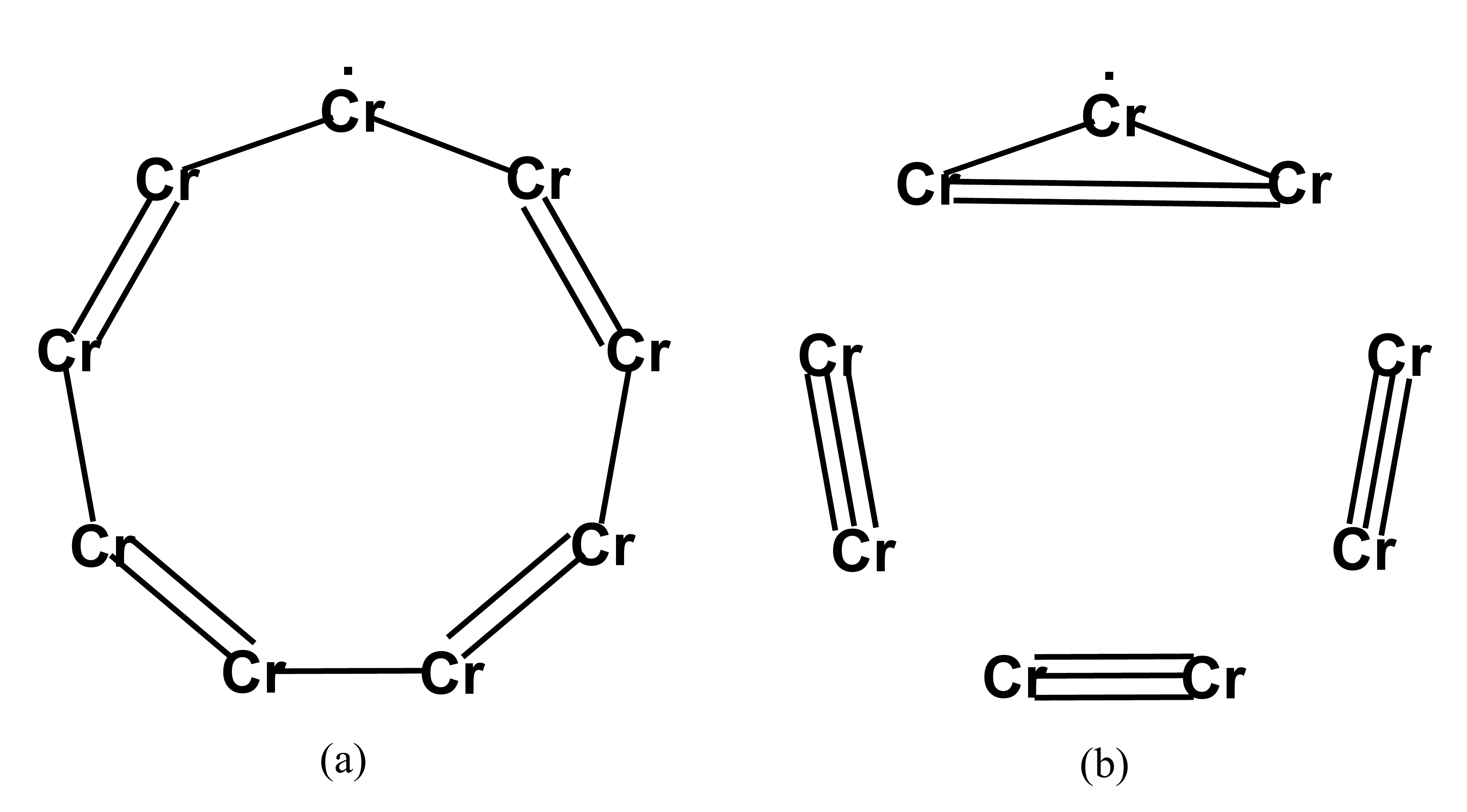}
\caption{\label{fig:cr9lewis}
Possible Lewis structures for the $S=1/2$ state of the [\ce{Cr9}] SMM.
Every Cr-Cr bond consists of d-orbitals and each Cr atom has three d-orbitals which yields an active space of (27e, 27o).
}
\end{figure}

It is interesting that the Cr atoms in the molecule are nearly in $\text{D}_\text{9h}$ symmetry. For $S=1/2$, this particular geometry leaves 9 different choices of the location of a unpaired electron. We tried a couple of different Lewis structures (or PP references) that have a unpaired electron on different Cr's, and CCVB methodologies all yielded very similar energies. Therefore, we picked Cr(1) (see Figure \ref{fig:cr9} for the label) to be the radical site for simplicity. We tried two different pairing schemes in CCVB. One of them is simply alternating single and double bonds for $S=1/2$ as shown in Figure \ref{fig:cr9lewis} (a). The other one forms a triangle among Cr(1), Cr(2), and Cr(9) while having triple-bonded $\ce{Cr2}$ for the rest as in Figure \ref{fig:cr9lewis} (b).

The UHF (or BS-DFT) approach \insertnew{can correctly separate} on $M_S = 3/2, 9/2, 15/2, 21/2, 27/2$, while states with other than those $M_S$ values will yield erroneously high energies. To obtain reasonable energies from UHF, we need to keep three electrons on each Cr to be the same spin.
As we have seen from the previous examples, CCVB can be applied to a much broader range of problems than UHF, and CCVB+i3 can be applied to even broader of strong correlation problems with SSC.

\begin{table}
  \centering
  \begin{tabular}{r|r|r|r|r|r|r}\hline
\multicolumn{1}{c}{$S$}\vline 
& \multicolumn{1}{c}{CCVB} \vline
& \multicolumn{1}{c}{CCVB+i3//CCVB} \vline
& \shortstack{HCI//CCVB\\($\epsilon_1=5\times10^{-4}$)}
& \shortstack{HCI//CCVB\\($\epsilon_1=10^{-4}$)}
& \shortstack{HCI//CCVB\\($\epsilon_1=10^{-5}$)}
& \shortstack{HCI//CCVB\\($\epsilon_1=10^{-6}$)}\\
\hline
1/2   &   45.82   &   1.05   &   63.81   &   0.67   &   0.65   &   0.64\\
3/2   &   0.36   &   0.34   &   517.42   &   0.07   &   0.07   &   0.06\\
5/2   &   0.33   &   0.31   &   530.89   &   0.06   &   0.05   &   0.05\\
7/2   &   0.32   &   0.31   &   63.20   &   0.05   &   0.06   &   0.05\\
9/2   &   0.31   &   0.28   &   505.25   &   0.06   &   0.05   &   0.05\\
11/2   &   0.31   &   0.28   &   0.16   &   0.06   &   0.05   &   0.04\\
13/2   &   0.29   &   0.25   &   0.13   &   0.04   &   0.04   &   0.04\\
15/2   &   0.28   &   0.22   &   0.09   &   0.04   &   0.05   &   0.04\\
17/2   &   0.25   &   0.23   &   0.08   &   0.04   &   0.04   &   0.03\\
19/2   &   0.17   &   0.13   &   0.05   &   0.04   &   0.06   &   0.04\\
21/2   &   0.13   &   0.12   &   0.03   &   0.02   &   0.02   &   0.02\\
23/2   &   0.05   &   0.05   &   0.02   &   0.01   &   0.01   &   0.01\\
25/2   &   0.07   &   0.07   &   0.01   &   0.00   &   0.00   &   0.00\\
27/2   &   0.00   &   0.00   &   0.00   &   0.00   &   0.00   &   0.00\\
\hline
  \end{tabular}
  \caption{
The relative energies (kcal/mol) of different spin states of \ce{Cr9}. The PP reference used for CCVB calculations corresponds to Figure \ref{fig:cr9lewis} (a).
CCVB+i3//CCVB denotes the CCVB+i3 energies evaluated with converged CCVB orbitals.
HCI//CCVB denotes the HCI energies evaluated with converged CCVB orbitals.
\revinsert{For $S=27/2$, every method is exact since ROHF is exact for that state.
The corresponding $S=27/2$ energy is -14101.38880 $E_h$.
These spin-gaps are directly comparable across different methods as they are measured with respect to this same energy.}
  }
  \label{tab:cr9energy1}
\end{table}

In Table \ref{tab:cr9energy1}, we see that HCI provides almost converged energies with $\epsilon_1 = 10^{-4}$. The HCI energies indicate that different spin states lie within 1 kcal/mol, which is the signature of SSC. Converging absolute energies up to the usual chemical accuracy, namely 1 kcal/mol, may not be appropriate to resolve the energy scale of strongly spin-correlated systems like this SMM system. CCVB and CCVB+i3 show more or less the same results except for the $S=1/2$ state. For the doublet state, there is a roughly 44 kcal/mol energy lowering going from 2P to 3P. This shows the significance of 3P substitutions when describing low-spin states of spin-frustrated systems.
As the CCVB+i3//CCVB results are all above the converged HCI//CCVB energies, CCVB+i3//CCVB is {\it practically} variational. 

Increasing the value of $\epsilon_1$ to $5\times10^{-4}$ results in catastrophic HCI failures for low-spin states, $S=1/2-9/2$. These energies are considered qualitatively wrong as the relevant energy scale is less than a kcal/mol in this system. 
In passing we note that the quality of CCVB orbitals for $S=1/2$ may be poor compared to other states given that there is a 0.60 kcal/mol energy jump going from $S=3/2$ to $S=1/2$ in HCI//CCVB. The unconverged HCISCF calculations indicate that spin-gaps \revinsert{are} of the order of 0.01 kcal/mol for all states.

\begin{table}
  \centering
  \begin{tabular}{r|r|r|r|r|r|r}\hline
\multicolumn{1}{c}{$S$}\vline 
& \multicolumn{1}{c}{CCVB} \vline
& \multicolumn{1}{c}{CCVB+i3//CCVB} \vline
& \shortstack{HCI//CCVB\\($\epsilon_1=5\times10^{-4}$)}
& \shortstack{HCI//CCVB\\($\epsilon_1=10^{-4}$)}
& \shortstack{HCI//CCVB\\($\epsilon_1=10^{-5}$)}
& \shortstack{HCI//CCVB\\($\epsilon_1=10^{-6}$)}\\
\hline
1/2  &  27.62  &  0.57  &  316.40  &  0.28  &  0.26  &  0.23\\
3/2  &  20.95  &  0.46  &  0.22  &  0.20  &  0.19  &  0.17\\
5/2  &  0.39  &  0.37  &  646.84  &  0.06  &  0.06  &  0.05\\
7/2  &  0.26  &  0.25  &  538.96  &  63.37  &  0.04  &  0.03\\
9/2  &  0.43  &  0.41  &  189.54  &  63.06  &  0.05  &  0.05\\
11/2  &  0.22  &  0.20  &  584.93  &  0.06  &  0.04  &  0.03\\
13/2  &  0.21  &  0.16  &  189.34  &  0.09  &  0.04  &  0.02\\
15/2  &  0.26  &  0.25  &  63.44  &  63.05  &  0.04  &  0.04\\
17/2  &  0.18  &  0.14  &  0.14  &  0.04  &  0.04  &  0.02\\
19/2  &  0.16  &  0.16  &  0.10  &  0.03  &  0.03  &  0.02\\
21/2  &  0.07  &  0.07  &  0.05  &  0.02  &  0.02  &  0.01\\
23/2  &  0.11  &  0.10  &  0.03  &  0.02  &  0.01  &  0.01\\
25/2  &  0.06  &  0.06  &  0.01  &  0.00  &  0.00  &  0.00\\
27/2  &  0.00  &  0.00  &  0.00  &  0.00  &  0.00  &  0.00\\
\hline
  \end{tabular}
  \caption{
  Same as Table \ref{tab:cr9energy1} except that the PP reference used here corresponds to Figure \ref{fig:cr9lewis} (b).
  }
  \label{tab:cr9energy2}
\end{table}

Table \ref{tab:cr9energy2} presents the solutions obtained using the PP reference in Figure \ref{fig:cr9lewis} (b). As CCVB energies are not invariant to the choice of a PP reference, we obtain different results. CCVB has a strong dependence on the PP reference because with this new reference the $S=1/2$ energy is 18 kcal/mol lower and the $S=3/2$ energy is 20 kcal/mol higher compared to the previous case. The qualitative failure of CCVB for $S=1/2$ and $S=3/2$ can be understood similarly to the D$_\text{3h}$ cases discussed above as we have a localized triangle in the PP reference. HCI//CCVB exhibits a catastrophic behavior for large $\epsilon_1$ values, but is adquately converged with $\epsilon_1 = 10^{-5}$. In the case of $\epsilon_1 = 5\times10^{-4}$, HCI//CCVB fails for all spin states lower than $S=15/2$ but $S=3/2$. 

\begin{table}
  \centering
  \begin{tabular}{r|r|r|r|r|r}\hline
\multicolumn{1}{c}{$S$}\vline 
& \multicolumn{1}{c}{CCVB+i3} \vline
& \shortstack{HCI//CCVB(a)\\($\epsilon_1=5\times10^{-4}$)}
& \shortstack{HCI//CCVB(a)\\($\epsilon_1=10^{-4}$)}
& \shortstack{HCI//CCVB(b)\\($\epsilon_1=5\times10^{-4}$)}
& \shortstack{HCI//CCVB(b)\\($\epsilon_1=10^{-4}$)}\\
\hline
1/2   &   104   &   8361   &   53061   &   17446   &   32707 \\
3/2   &   114   &   9557   &   31082   &   12661   &   42455 \\
5/2   &   121   &   5793   &   14178   &   12291   &   36431 \\
7/2   &   125   &   6678   &   23750   &   9217   &   49014 \\
9/2   &   126   &   4828   &   15764   &   7936   &   35872 \\
11/2   &   124   &   2797   &   10107   &   4266   &   28343 \\
13/2   &   119   &   1768   &   6029   &   6851   &   30096 \\
15/2   &   111   &   1287   &   4743   &   2149   &   13229 \\
17/2   &   100   &   912   &   3039   &   2320   &   8682 \\
19/2   &   86   &   484   &   1218   &   1418   &   5804 \\
21/2   &   69   &   310   &   781   &   860   &   1715 \\
23/2   &   49   &   98   &   212   &   186   &   371 \\
25/2   &   26   &   52   &   87   &   56   &   76 \\
27/2   &   0   &   0   &   0   &   0   &   0 \\
\hline
  \end{tabular}
  \caption{
The number of independent wavefunction parameters used in each method in the [\ce{Cr9}] SMM. CCVB(a) and CCVB(b) denote the CCVB orbitals with the PP references in Figure \ref{fig:cr9lewis} (a) and (b), respectively.
  }
  \label{tab:cr9params}
\end{table}

Based on the results discussed in Table \ref{tab:cr9energy1} and Table \ref{tab:cr9energy2}, we conclude that CCVB+i3 is less sensitive to the underlying PP reference and its accuracy lies somewhere in between HCI of $\epsilon_1 = 10^{-4}$ and $\epsilon_1 = 5\times10^{-5}$. We further compare those two methodologies in terms of the number of independent wavefunction parameters and we emphasize the compactness of the CCVB+i3 wavefunction as shown in Table \ref{tab:cr9params}. 
\insertnew{Due to the larger system}, the difference in the number of parameters is larger in [\ce{Cr9}] than in \ce{P5} (shown in Table \ref{tab:p5params}). CCVB+i3 has a 300-500 times smaller number of parameters than HCI ($\epsilon_1=10^{-4}$) for $S=1/2$ in [\ce{Cr9}] whereas in \ce{P5} it was only 150 times smaller compared to HCISCF ($\epsilon_1 = 10^{-4}$). One may think that orbital optimization must help reduce the number of determinants in HCI in the case of [\ce{Cr9}]. However, as the effect of orbital optimization is very small in this system we believe that the conclusion here will not be altered.

The situation will become only more favorable to CCVB when studying \insertnew{larger molecules or} bulk materials as there will be too many determinants to include for HCISCF even to just achieve a similar accuracy as CCVB. The strength of CCVB is at the use of a CC-type expansion to avoid including an exponential number of wavefunction parameters while being able to describe strong spin correlation and yielding a size-consistent, spin-pure energy and wavefunction.

Lastly, we note that the singular Jacobian problem we faced in this system is not necessarily an indication of redundant wavefunction parameters. When evaluated with a solution to CC amplitude equations, a CC Jacobian is often interpreted as an equation-of-motion (EOM) CC Hamiltonian. Eigenvalues of the CC Jacobian are excitation gaps. We checked the first 5 roots of each spin state from HCI//CCVB and observed that the first 5 roots are all within 10 $\mu$H. This is consistent with the singular values of Jacobian we observed in CCVB+i3. Strongly spin-correlated systems have a dense spectrum of low-lying excited states which would necessarily imply (nearly) singular CC Jacobians. It is thus important to develop a better amplitude solver to tackle strongly correlated systems with non-linear CC wavefunctions. In passing we note that this may indicate that the EOM treatment to CCVB can yield quite accurate excitation gaps for strongly correlated systems. This will be further investigated along with EOM-CCVB-SD in future work.
\section{Conclusions and Outlook}
In this paper, we tested the CCVB \revreplace{ans{\"a}tz}{ansatz} on spin-frustrated systems with SSC. Those systems include \ce{N3}, \ce{V3O3}, the cubane subunit of oxygen-evolving complex, \ce{P5}, and the [\ce{Cr9}] single molecular magnet. We showed that the model catastrophically fails to describe the lowest-spin states of such systems. As an attempt to fix this problem, we introduced an improved electron correlation model, CCVB+i3, which includes 3-pair correlations that are missing in CCVB. Our working hypothesis is that the new model can in principle reach any bond dissociation limits exactly within an active space, and we numerically showed that it provides a qualitatively correct description of those spin-frustrated systems when CCVB fails. It was also emphasized that the new model involves the same number of independent wavefunction parameters as CCVB and scales the same.

We compared CCVB+i3 against an exponential-scaling heat-bath CI (HCI) method for \ce{P5} and [\ce{Cr9}]. For those systems, HCI was able to converge the energy below 0.1 kcal/mol with a reasonable amount of computational work. CCVB+i3 energies are 1-2 kcal/mol and 0.5 kcal/mol above those of HCI in the case of \ce{P5} and [\ce{Cr9}], respectively. We emphasized the promise of CCVB+i3 by comparing the number of independent wavefunction parameters against HCI. To achieve a similar accuracy, HCI involves roughly a 300-500 times larger number of parameters than CCVB+i3 in [\ce{Cr9}]. Towards the application to \insertnew{large molecules and} bulk strongly correlated systems, this scaling will become only more favorable to CCVB.

There are many promising future developments of CCVB and CCVB+i3. The most interesting extension is perhaps to incorporate missing dynamic correlations not only within the active space but also outside the active space. This could be achieved either using density functional theory, \cite{Grimme1999,Kurzweil2009,Gagliardi2017} perturbation theory,\cite{Kucharski1988,Wolinski1989,Andersson1990,Murphy1991,Hirao1992,Dyall1995,Beran2006,Xu2013,Evangelista2014,Li2015,Lee2018} or extended random phase approximations.\cite{Pernal2014,Pastorczak2015} Another interesting extension is to implement nuclear gradients and other properties of CCVB. In particular, nuclear gradients will be particularly useful as CCVB equilibrium geometries are quite close to CASSCF at least in small molecular systems that have been studied. Those two developments will help to put CCVB among the set of routinely applicable electron correlation models.

Other theoretical questions of CCVB include whether it is necessary to go beyond the IAA treatment of the 3P substitutions. The scope of CCVB+i3 remains unclear although the numerical results so far indicate that it is capable of dissociating any number of bonds. We are investigating its relation to the spin-projected generalized Hartree-Fock (SGHF) wavefunction to learn more about its scope. The full CCVB-3 model is not size-consistent as shown in this work. The next level of a size-consistent CCVB method would then be the one that includes everything up to the 4-pair contributions in the 3-pair amplitude equation. This will be an interesting model to explore although the overall cost will no longer be the same as CCVB. 

Another question is then whether we can take either CCVB+i3 or more sophisticated wavefunctions and generalize them to a full CC model with singles, doubles and triples. Given the \insertnew{promising initial} success of CCVB-SD which generalizes CCVB, it will be interesting to generalize a variant of the 3-pair model to a full CC model as well.
\insertnew{The main strength of CCVB+i3, as well as CCVB, is for systems where the strong correlations are primarily strong spin correlations. Systems where strong charge fluctuations are also important (e.g. mixed valence metal ions) require ionic excitations that are excluded in CCVB and CCVB+i3. They are restored in full CC model such as CCVB-SD.}
As mentioned in the main text, we are also investigating excited states of CCVB and CCVB-SD within the equation of motion framework.
%
\section{Supplementary Material}
The supplemental material of this work is available online which includes the proof of Eq. \eqref{eq:zeroquad}, Eq. \eqref{eq:T3} in terms of computable quantities and the CCVB+i3 Jacobian, Lagrangian and associated derivatives for optimization.
\section{Acknowledgement}
We thank Paul R. Horn for implementing various generic solvers in $\texttt{Q-Chem}$ that are used in this work, J{\'e}r{\^o}me Gonthier for helping some of the crosscheck calculations performed with $\texttt{Psi4}$, James E. T. Smith and Sandeep Sharma for providing an early access to $\texttt{Dice}$ used in this work, and Brett Van Der Goetz and Eric Neuscamman for helpful discussions on the singular Jacobian problem.
J. L. also thanks Soojin Lee for consistent encouragement and support.
This work was supported by the Director, Office of Science, Office of Basic Energy Sciences, of the U.S. Department of Energy under Contract No.~DE-AC02-05CH11231.

\bibliography{iepa3p_ms_rev1}
\bibliographystyle{achemso}
\end{document}